\documentclass[twocolumn]{article}  
\usepackage{graphicx}
\usepackage[T1]{fontenc}
\usepackage[subrefformat=parens]{subcaption}

\makeatletter
\let\cl@chapter\undefined
\makeatletter

\usepackage{amsmath,amssymb}   
\usepackage{amsthm} 
\usepackage{mathtools} 
\usepackage[numbers,sort&compress]{natbib}
\usepackage[margin=15mm]{geometry}
\usepackage{hyperref}
\hypersetup{colorlinks=true,pdfborder={0 0 0}}
\usepackage{cleveref}  
\usepackage{amsbsy}
\usepackage{bm}
\usepackage{url}
\usepackage{color}
\usepackage{multirow}
\usepackage{tabularx}
\usepackage{booktabs}
\usepackage{autobreak}
\allowdisplaybreaks 

\newcommand{\inlineTag}{
    \refstepcounter{equation}
    \bgroup\normalfont\normalcolor (\theequation)\egroup}

\crefname{equation}{Eq.}{Eqs.}%
\crefname{figure}{Fig.}{Figs.}%

\usepackage{algorithm}
\usepackage[noend]{algpseudocode}

\algnewcommand{\Break}{\textbf{break}}
\algrenewcommand\algorithmicindent{1em}

\everydisplay{\small} 

\newcommand\Erase{\bgroup\markoverwith{\textcolor{red}{\rule[.5ex]{2pt}{0.4pt}}}\ULon}

\begin{document}

    \title{
        Probabilistic Prediction of Ship Maneuvering Motion using Ensemble Learning with Feedforward Neural Networks
    }
    
    \author{
        Kouki Wakita$^{1}$ \and
        Youhei Akimoto$^{2,3}$ \and
        Atsuo Maki$^{1}$
    }

    \date{%
    \flushleft{\footnotesize
        $^1$Osaka University, 2-1 Yamadaoka, Suita, Osaka, Japan \\%
        $^2$Faculty of Engineering, Information and Systems, University of Tsukuba, 1-1-1 Tennodai, Tsukuba, Ibaraki 305-8573, Japan \\
        $^3$RIKEN Center for Advanced Intelligence Project, 1-4-1 Nihonbashi, Chuo-ku, Tokyo 103-0027, Japan \\[2ex]%
        Email: kouki\_wakita@naoe.eng.osaka-u.ac.jp; maki@naoe.eng.osaka-u.ac.jp \\
    }}


    \maketitle

    \begin{abstract}
        In the field of Maritime Autonomous Surface Ships (MASS), the accurate modeling of ship maneuvering motion for harbor maneuvers is a crucial technology. Non-parametric system identification (SI) methods, which do not require prior knowledge of the target ship, have the potential to produce accurate maneuvering models using observed data. However, the modeling accuracy significantly depends on the distribution of the available data. To address these issues, we propose a probabilistic prediction method of maneuvering motion that incorporates ensemble learning into a non-parametric SI using feedforward neural networks. This approach captures the epistemic uncertainty caused by insufficient or unevenly distributed data. In this paper, we show the prediction accuracy and uncertainty prediction results for various unknown scenarios, including port navigation, zigzag, turning, and random control maneuvers, assuming that only port navigation data is available. Furthermore, this paper demonstrates the utility of the proposed method as a maneuvering simulator for assessing heading-keeping PD control. As a result, it was confirmed that the proposed method can achieve high accuracy if training data with similar state distributions is provided, and that it can also predict high uncertainty for states that deviate from the training data distribution. In the performance evaluation of PD control, it was confirmed that considering worst-case scenarios reduces the possibility of overestimating performance compared to the true system. Finally, we show the results of applying the proposed method to full-scale ship data, demonstrating its applicability to full-scale ships.
                    \end{abstract}

    \section{Introduction}\label{sec:intro}
        A simulator for ship maneuvering motions that allows for the simple and safe evaluation of ship maneuvering systems is a practical and valuable tool, as it has the potential to reduce the need for tests using actual ships. Simulation methods using system-based mathematical models for ships (hereafter referred to as ``maneuvering models'') are widely utilized because they can simulate ship maneuvers with small computational resources. Thus, a methodology for constructing maneuvering models capable of simulating ship maneuvering motions with high accuracy holds significant practical value. Therefore, a methodology for obtaining a maneuvering model that enables accurate prediction has practical value.
        
        In recent years, research and development of Maritime Autonomous Surface Ships (MASS) has been actively promoted, and the automation of harbor maneuvers is one of the most important issues in the realization of MASS. The term ``harbor maneuvers'' in this paper refers to operations within the harbor, including entering and leaving the port, approaching the berth, berthing, and unberthing. For this reason, it is expected that there will be an increasing demand for maneuvering models for harbor maneuvers in the future.
    
        The maneuvering model for harbor maneuvers needs to be able to simulate not only standard maneuvers with relatively high and stable forward speeds, but also maneuvers at low speeds. Therefore, the range of maneuvering motions that required to be modeled is larger. Low-speed maneuvers, such as berthing and unberthing, may require special maneuvers, such as crabbing, rotation, and moving astern, and require more frequent actuator operations\cite{Miyauchi2022data,mwange2024}. Therefore, the creation of a highly accurate maneuvering model for harbor maneuvers is a difficult task.
            
        \subsection{Related works}\label{subsec:works}
            Many ship maneuvering models are constructed based on hydrodynamic principles. For instance, the MMG model, proposed by the research group of the Japan Towing Tank Conference, is a mathematical model composed of submodels that represent hydrodynamic forces induced by the hull, propeller, rudder, and other actuators \cite{Ogawa1978}. The MMM model is modular, so it has the advantage of allowing partial design modifications by adjusting specific sub-models during the shipbuilding process. Extensive research has also been conducted on the MMG model for low-speed maneuvering \cite{Kose1984,Fujino1990,KOBAYASHI1994,ISHIBASHI1996,Yoshimura2009}.
    
            Most hydrodynamic forces can be measured through captive model tests using scale model ships, and planar motion mechanism (PMM) tests \cite{Chislett1965} and circular motion tests (CMT) \cite{Koyama1975} are frequently employed. Based on the measured forces, it is possible to determine the maneuvering model and its parameters. It is also possible to derive specific parameters using empirical formulas; for example, Motora’s chart \cite{Motora1959En} is commonly used for added mass and added moment of inertia, while some coefficients of hull hydrodynamic forces can be obtained using Inoue's formula \cite{Inoue1981}, Kijima's formula \cite{Kijima1990}, and Yoshimura's formula \cite{Yoshimura2009}.
    
            As efforts are made to more accurately represent hydrodynamic forces, the complexity of the model increases. Even when empirical formulas and previously accumulated captive model test data are available, it is not always possible to determine all the parameters. When conducting captive model tests, specialized facilities such as towing tanks or maneuvering basins, along with the expertise to properly operate them, are required. Furthermore, since the forces measurable on full-scale ships are usually limited, it is necessary to appropriately scale the results of model-scale tests. However, this process may be challenged by scale effects.
    
            As another approach, the methodology of estimating maneuvering models using system identification (SI) has long been studied \cite{ASTROM1976,Abkowitz1980,KALLSTROM1981,YOON2003,Sutulo2014,Sawada2021,Miyauchi2022,Sutulo2023}. SI does not necessarily require direct measurement of forces; instead, it allows modeling based on time-series data of kinematic variables and control inputs. This makes it possible to apply it to full-size ships.

            In several literature \cite{HE2022,SHEN2023}, SI for maneuvering models are classified into parametric SI and non-parametric SI. Parametric SI utilizes prior knowledge about the target system, which is often incorporated into a maneuvering model. Examples of such models include the MMG model \cite{Miyauchi2022,CHEN2022,Sutulo2023,SUYAMA2024} and the Abkowitz model \cite{Abkowitz1980,Luo2009,Zhang2011,Miyauchi2023}. In parametric SI, if the prior knowledge enables a simple and efficient representation of the system, it is possible to estimate the parameters using relatively small amounts of data. However, if the system is not appropriately represented, the estimation accuracy may deteriorate.
            
            In the context of ship maneuvering models, the application of the Extended Kalman Filter (EKF) has been studied from early on \cite{Abkowitz1980,KALLSTROM1981,ASTROM1976,YOON2003}, and there has also been a lot of research into the application of regression analysis techniques such as Support Vector Regression (SVR) \cite{Luo2009,Luo2014,ZHU2019,WANG2019}. If the identification process does not need to be carried out online, offline identification, which allows access to all measurement data at all times, is more effective than online identification \cite{Sutulo2014}. In offline identification, model parameters can be determined by optimizing the error between the observed and simulated data. These errors can be evaluated using metrics such as the sum of squared errors \cite{Miyauchi2022}, the Hausdorff metric \cite{Sutulo2014}, or the negative log-likelihood related to observation noise \cite{Wakita2024}. Furthermore, offline identification enables the use of global optimization algorithms, such as genetic algorithms (GA) \cite{Sutulo2014,Sutulo2023} and covariance matrix adaptation evolution strategy (CMA-ES) \cite{Miyauchi2022,Miyauchi2023,SUYAMA2024}.

            Non-parametric SI is an SI that does not rely on prior knowledge of the target system, and often uses machine learning-based surrogate models such as artificial neural networks (ANN) and support vector machines (SVM) that use kernel functions. Since non-parametric SI does not require a priori determination of the model structure, it can be applied to any type of actuator configuration, and has the advantage that it can be used by users who are not hydrodynamics experts. Furthermore, machine learning surrogate models excel in representing nonlinear functions. For instance, ANN has strong function approximation capabilities \cite{Cybenko1989,HORNIK1991251}, and numerous studies \cite{MOREIRA2003,Moreira2012,RAJESH2008,Zhang2013,HE2022} have demonstrated its ability to effectively capture the nonlinearities of maneuvering models. Additionally, recurrent neural networks (RNN) \cite{OSKIN2013,Hao2022,Wakita2022} and long short-term memory (LSTM) networks \cite{WOO2018,Jiang2022} can be employed to account historical effects. SVM and Gaussian process (GP) models \cite{ARIZARAMIREZ2018,Xue2020,XUE2022} leverage the kernel trick to capture nonlinear patterns in maneuvering models with relatively low computational cost.
    
            Non-parametric SI, lacking hydrodynamic background, is highly dependent on the distribution and quality of the available data. Unfortunately, since harbor maneuvers includes both standard maneuvers and low-speed maneuvers, a wide range data of maneuvering motions is required. On the other hand, conducting full-scale ship trials incurs high costs, and it is not always feasible to collect large amounts of data. For existing ships, daily operational data may be available. These operational data tends to be unevenly distributed. Thus, it is impractical to collect maneuvering motion data that covers all feasible motions during harbor maneuvers. As a result, non-parametric SI can not always provide a accurate maneuvering model for harbor maneuvers.
        
        \subsection{Objective of this study}\label{subsec:object}
            Discrepancies between simulation and physical environments can result in unexpected performance degradation in the development of maneuvering systems. In particular, in model predictive control (MPC), there is a possibility that model errors will be actively misused. In harbor maneuvers, where the distance between the ship and obstacles is relatively short, performance degradation of the maneuvering system could result in collisions or allisions, potentially causing fatal human and economic damages. Thus, in developing maneuvering systems for safety-critical tasks such as harbor maneuvers, it is essential either to use highly accurate simulators, or a simulator that can anticipate the possibility of performance deterioration in advance. This approach reduces the possibility of overestimating maneuvering systems in the simulation.
    
            However, non-parametric SI does not always provide a highly accurate maneuvering model for the entire operational range of a maneuvering system. In particular, when the state transitions outside the distribution of the given dataset, the prediction error is likely to increase significantly. Therefore, it is essential for maneuvering models obtained through non-parametric SI to incorporate the ability to provide the epistemic uncertainty caused by insufficient or unevenly distributed data.
            
            Several studies on non-parametric SI for maneuvering model \cite{ARIZARAMIREZ2018,Xue2020,Xue2021} have proposed models that provide prediction reliability by using Gaussian processes (GP). However, these studies primarily focus on uncertainties arising from external disturbances such as ocean currents, waves, and wind. To the best of our knowledge, no research has explored the epistemic uncertainties of maneuvering models.
    
            In recent years, numerous studies have been conducted on understanding and quantifying the predictive uncertainty of ANN \cite{Gawlikowski2023}. Bayesian inference is one approach that can capture epistemic uncertainty. However, exact Bayesian inference for neural networks requires substantial computational resources, limiting its application to problems with smaller dimensions and datasets. One method for approximating Bayesian inference with fewer computational resources is the ensemble method. This approach has been shown to scale linearly in terms of memory and computational requirements with the number of ensemble members \cite{Lakshminarayanan2017,rahaman2021}.
    
            The objective of this paper is to address the epistemic uncertainty by incorporating Ensemble Learning into a Non-parametric SI using ANNs for predicting ship maneuvering motions. This paper proposes an Ensemble Learning approach for ANNs and a probabilistic motion prediction method based on trained models. The proposed method allows for highly accurate predictions for states with abundant data and highly uncertain predictions for states with little or no available data.
            The contributions of this paper are as follows: 
            \begin{itemize}
                \item We propose Ensemble Learning methods based on the Non-parametric SI method using ANNs in previous studies \cite{Wakita2022, Wakita2024}.
                \item To improve the fitting accuracy of kinematic variables in maneuvering simulations based on a maneuvering model, we introduce an ANN for initial state estimation into the SI method for the maneuvering model proposed in previous research \cite{Wakita2022, Wakita2024}. 
                \item We compare the two trajectory sampling methods proposed by Chua et al. \cite{Chua2018} and propose a probabilistic prediction method of ship maneuvering motions that can efficiently capture epistemic uncertainty.
                \item As an example of applying the proposed method to the assessment of maneuvering systems and control algorithms, We show the experiment results of gain evaluation in heading-keeping PD control, and demonstrate the effectiveness of the simulator.
            \end{itemize}
            This paper presents the validation results of the proposed approach using numerical simulation data from a model ship and operational data from a full-scale ship. The simulation data include port navigation \cite{Miyauchi2022data, mwange2024}, zigzag maneuvers, turning, random control maneuvers. The relationship between prediction accuracy and the predicted uncertainty is evaluated using this dataset.
    
            The rest of this paper is organized as follows: \Cref{sec:pre} explains preliminaries, \Cref{sec:learning} details the proposed ensemble learning method, \Cref{sec:prediction} describes the probabilistic prediction for maneuvering motions, \Cref{sec:sim_exp,sec:nav_exp} presents experiment results, \Cref{sec:discussion} discusses findings, and \Cref{sec:conclusion} concludes this paper.
            
    \section{Preliminaries}\label{sec:pre}
        \subsection{Notation}\label{sec:note}
            The symbol $\mathbb{R}$ denotes the set of real numbers, and $\mathbb{R}^n$ represents the $n$-dimensional Euclidean space. Let $\mathbb{S} = [0, 2\pi]$ denote the set of angles. If we define a real vector as $\boldsymbol{x} \in \mathbb{R}^n$ and a real matrix as $\boldsymbol{A} \in \mathbb{R}^{n \times n}$, $\|\boldsymbol{x}\|$ denotes the Euclidean norm of $\boldsymbol{x}$, defined as $(\boldsymbol{x}^{\mathsf{T}}\boldsymbol{x})^{1/2}$. Similarly, $\|\boldsymbol{x}\|_{\boldsymbol{A}}$ denotes the Euclidean norm of $\boldsymbol{x}$ weighted by $\boldsymbol{A}$, defined by $(\boldsymbol{x}^{\mathsf{T}}\boldsymbol{A}\boldsymbol{x})^{1/2}$. 
            
        \subsection{Subject ship}\label{subsec:subjectship}
            \begin{figure}[t]
                \centering
                \includegraphics[width=\linewidth]{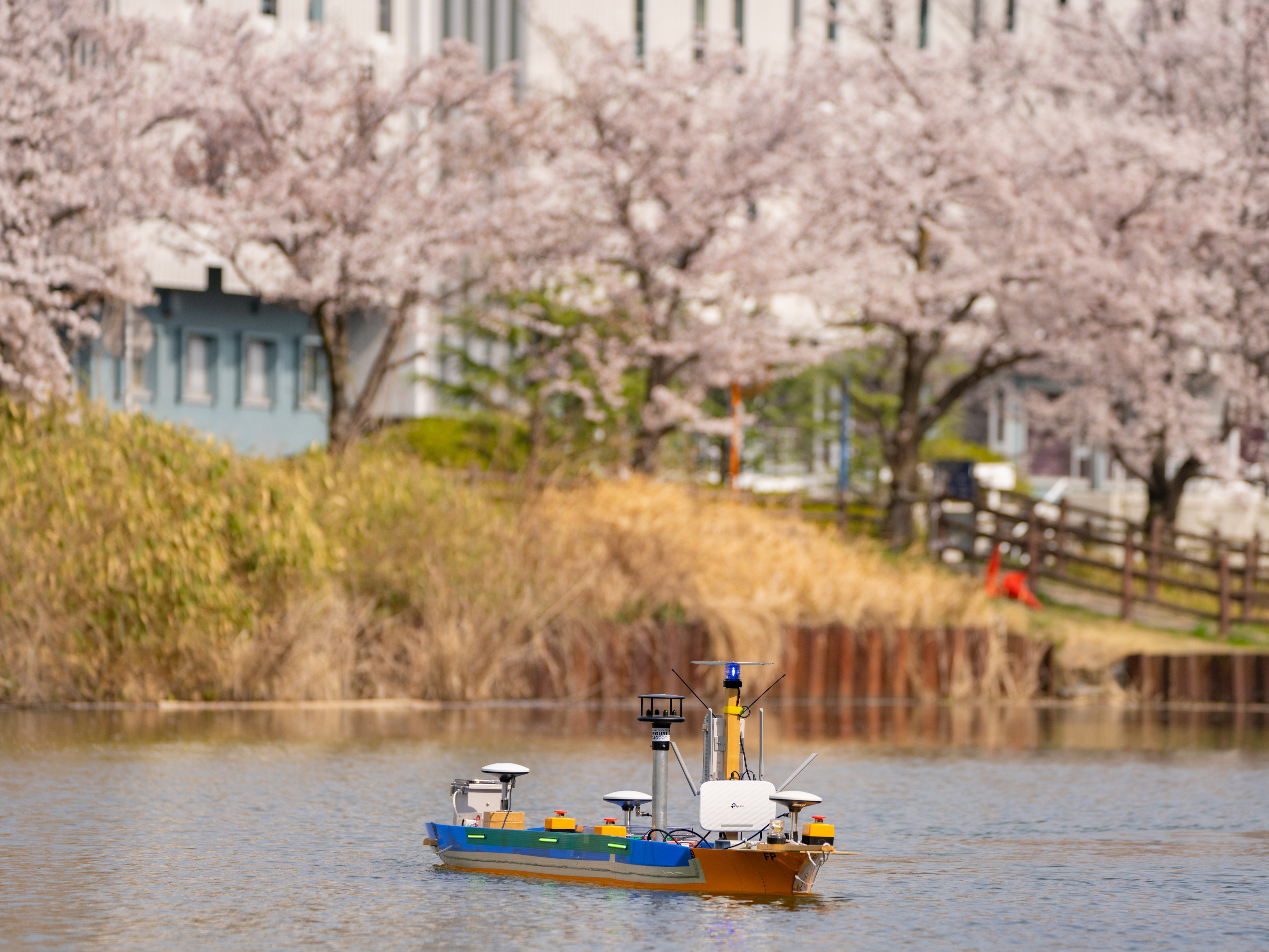}
                \caption{A subject model ship.}
                \label{fig:subjectship}
            \end{figure}
            
            \begin{table}[t]
                \centering
                \caption{Limitations of the actuator state variables}
                \begin{tabularx}{\linewidth}{llX}\toprule
                    Symbols & Range & Description \\ \midrule
                    $\delta_{\mathrm{P}} \ (\mathrm{deg.})$ & $[-105, 35]$ & Port side rudder angle. \\
                    $\delta_{\mathrm{S}} \ (\mathrm{deg.})$ & $[-35, 105]$ & Starboard side rudder angle. \\
                    $n_{\mathrm{P}} \ (\mathrm{rps})$ & $\{10\}$ & Propeller revolution number, which is constant in this study. \\
                    $n_{\mathrm{BT}} \ (\mathrm{rps})$ & $[-30, 30]$ & Bow thruster revolution number. \\
                    \bottomrule
                \end{tabularx}
                \label{tab:controllimit}
            \end{table}
    
            The subject ships are a 3-meter model ship equipped with a single-propeller, a VecTwin rudder system and a bow thruster as shown in \Cref{fig:subjectship}, and a full-scale version of the same ship. VecTwin rudder system \cite{JAPANHAM} consists of two fishtail rudders positioned behind a single fixed pitch propeller, with each rudder operating independently to achieve high maneuverability \cite{Hasegawa2006}. This system enables the execution of specialized maneuvers, such as hovering, crabbing, astern, and rotation, while maintaining a constant forward propeller rotation, as long as at least one side thruster is available \cite{YASUKAWA2011, Rachman2023, Wakita2024MBRL}.
            The rudder angles of the port and starboard rudders, the propeller revolution speed, and the bow thruster revolution speed are represented by $\delta_{\mathrm{P}}$, $\delta_{\mathrm{S}}$, $n_{\mathrm{P}}$, and $n_{\mathrm{BT}}$, respectively. The upper and lower limits for each of these parameters are listed in \Cref{tab:controllimit}. In this study, since the propeller revolution speed is fixed, the actuator state vector is defined as $\boldsymbol{u} \equiv \left(\delta_{\mathrm{P}}, \delta_{\mathrm{S}}, n_{\mathrm{BT}}\right)^{\mathsf{T}}$.
        
        \subsection{Ship maneuvering motion on calm water}\label{subsec:motion}
            \begin{figure}[t]
                \centering
                \includegraphics[width=0.9\linewidth, trim=20 35 20 20, clip]{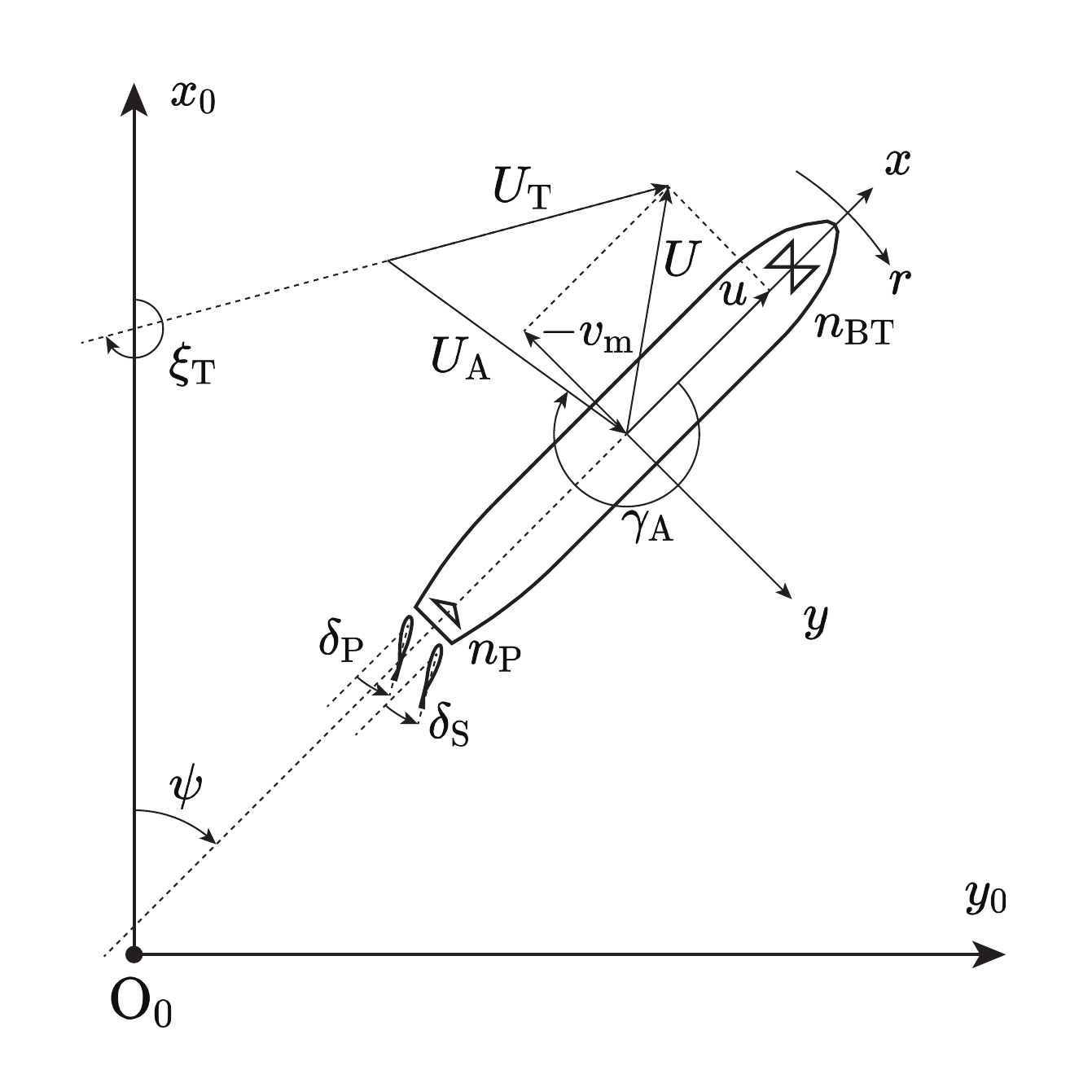}
                \caption{Coordinate systems.}
                \label{fig:coo}
            \end{figure}
            
            This study focuses on the three degrees of freedom maneuvering motion on calm water, considering wind disturbances, as illustrated in \Cref{fig:coo}. The ship's motion is expressed by the pose vector $\boldsymbol{\eta} \equiv \left( x_{0}, y_{0}, \psi \right)^{\mathsf{T}} \in \mathbb{R}^{2}\times\mathbb{S}$ and the velocity vector $\boldsymbol{\nu} \equiv \left( u, v_{\mathrm{m}}, r \right)^{\mathsf{T}} \in \mathbb{R}^{3}$. Here, $\left( x_{0}, y_{0} \right)$ represents the midship position in the earth-fixed coordinate system $\mathrm{O_{0}}$-$x_{0}y_{0}z_{0}$, $\psi$ is the yaw angle, $\left( u, v_{\mathrm{m}} \right)$ are surge and sway velocities, respectively, and $r$ represents the yaw rate. Additionally, the state vector of the ship is defined as $\boldsymbol{x} \equiv \left( \boldsymbol{\eta}^{\mathsf{T}}, \boldsymbol{\nu}^{\mathsf{T}} \right)^{\mathsf{T}} \in \mathbb{R}^{6}$.
            
            Moreover, forces induced by the wind are taken into account. Although wind speed and direction vary with time $t$, in this study, they are assumed to be spatially uniform. The true wind is represented as the vector $\boldsymbol{w}_{\mathrm{T}} \equiv \left( U_{\mathrm{T}}, \xi_{\mathrm{T}} \right)^{\mathsf{T}} \in \mathbb{R}\times\mathbb{S}$, where $U_{\mathrm{T}}$ and $\xi_{\mathrm{T}}$ denote the true wind speed and direction, respectively. The reference direction for $\xi_{\mathrm{T}}$ is set such that wind blows from the positive to the negative $x_{0}$ direction. However, the wind force acting on the ship is determined by the apparent wind, not the true wind. The apparent wind is noted by the vector $\boldsymbol{w}_{\mathrm{A}} \equiv \left( U_{\mathrm{A}}, \gamma_{\mathrm{A}} \right)^{\mathsf{T}} \in \mathbb{R}\times\mathbb{S}$, where $U_{\mathrm{A}}$ and $\gamma_{\mathrm{A}}$ represent the apparent wind speed and direction, respectively. The reference direction for $\gamma_{\mathrm{A}}$ is set such that the wind blows from the bow to the stern of the ship.
            
            To illustrate the relationship between true wind and apparent wind, we define the true and apparent wind velocity vector in the ship-fixed coordinate system as $\boldsymbol{w}_{\mathrm{T}}^{\prime} \equiv \left( U_{\mathrm{T}}\cos\left(\xi_{\mathrm{T}}-\psi\right), U_{\mathrm{T}}\sin\left(\xi_{\mathrm{T}}-\psi\right) \right)^{\mathsf{T}} \in \mathbb{R}^{2}$ and $\boldsymbol{w}_{\mathrm{A}}^{\prime} \equiv \left(U_{\mathrm{A}}\cos\gamma_{\mathrm{A}}, U_{\mathrm{A}}\sin\gamma_{\mathrm{A}}\right)^{\mathsf{T}} \in \mathbb{R}^{2}$, respectively, and the ship velocity vector is defined as $\boldsymbol{v} \equiv \left( u, v_{\mathrm{m}}\right)^{\mathsf{T}} \in \mathbb{R}^{2}$. 
            Then, the relationship between the true wind vector and the apparent wind vector can be expressed as:
            \begin{equation}
                \boldsymbol{w}_{\mathrm{A}}^{\prime} = \boldsymbol{w}_{\mathrm{T}}^{\prime} - \boldsymbol{v}
                \enspace.
                \label{eq:wTprime2wAprime}
            \end{equation}
            Therefore, given the surge and sway velocity $\left( u, v_{\mathrm{m}} \right)$ and yaw angle $\psi$, the apparent wind speed and apparent wind direction can be calculated as follows:
            \begin{equation}
                \left\{
                \begin{aligned}
                    U_{\mathrm{A}} &= \sqrt{ \left( U_{\mathrm{T}}\cos\left(\xi_{\mathrm{T}}-\psi\right) - u \right)^{2} + \left( U_{\mathrm{T}}\sin\left(\xi_{\mathrm{T}}-\psi\right) - v_{\mathrm{m}} \right)^{2} }\\ 
                    \gamma_{\mathrm{A}} &= \mathrm{arctan2}\left(U_{\mathrm{T}}\sin\left(\xi_{\mathrm{T}}-\psi\right) - v_{\mathrm{m}}, U_{\mathrm{T}}\cos\left(\xi_{\mathrm{T}}-\psi\right) - u \right)
                    \enspace.
                \end{aligned}
                \right.
                \label{eq:wT2wA}
            \end{equation}
            
            The kinematics and dynamics of the ship are assumed to be sufficiently represented by the following system of ordinary differential equations:
            \begin{subequations} \label{eq:kinamatics_and_dynamics}
                \begin{align}
                \dot{\boldsymbol{\eta}} &= \boldsymbol{R}\left(\boldsymbol{\eta}\right) \boldsymbol{\nu}
                \label{eq:kinamatics} \\
                \dot{\boldsymbol{\nu}} &= \boldsymbol{F}\left(\boldsymbol{\nu}, \boldsymbol{u}, \boldsymbol{w}_{\mathrm{A}}\right) \enspace,
                \label{eq:dynamics} 
                \end{align}
            \end{subequations}
            where $\boldsymbol{R}\left(\boldsymbol{\eta}\right) \in \mathbb{\boldsymbol{R}}^{3\times3}$ is the rotation matrix from the $\mathrm{O}$-$xyz$ frame to the $\mathrm{O}_{0}$-$x_{0}y_{0}z_{0}$ frame, and $\boldsymbol{F}$ is an unknown vector function representing the maneuvering model. The rotation matrix $\boldsymbol{R}\left(\boldsymbol{\eta}\right)$ is defined by:
            \begin{equation}
                \boldsymbol{R}\left(\boldsymbol{\eta}\right) =
                \begin{bmatrix}
                    \cos \psi & -\sin \psi & 0 \\
                    \sin \psi & \cos \psi & 0 \\
                    0 & 0 & 1 \\
                \end{bmatrix}
                \enspace.
                \label{eq:rotation_matrix}
            \end{equation}
            For simplicity of notation, the composite function of \Cref{eq:dynamics,eq:wT2wA} is expressed as:
            \begin{equation}
                \dot{\boldsymbol{\nu}} =\boldsymbol{F}^{\prime}\left(\boldsymbol{x}, \boldsymbol{u}, \boldsymbol{w}_{\mathrm{T}}\right) \enspace.
                \label{eq:wT2wA_dynamics}
            \end{equation}
            Furthermore, \Cref{eq:kinamatics,eq:wT2wA_dynamics} is collectively expressed as:
            \begin{equation}
                \dot{\boldsymbol{x}} =\boldsymbol{f}\left(\boldsymbol{x}, \boldsymbol{u}, \boldsymbol{w}_{\mathrm{T}}\right) \enspace.
                \label{eq:concat_kinamatics_and_dynamics}
            \end{equation}
            
        \subsection{Feedforward Neural network}\label{subsec:fnn}
            A feedforward neural network (FNN) is a standard ANN model that is frequently used in machine learning tasks. FNNs have high approximation capabilities \cite{Cybenko1989,HORNIK1991251} and are capable of learning from large datasets through mini-batch training \cite{Krizhevsky2012,Hinton2012}, making them a highly useful and powerful tool. In this section, we describe the FNN used in this paper, which is composed of fully connected layers and activation functions.  
            
            A fully connected layer that transforms $N_{\mathrm{NN}}$-dimensional variables into $M_{\mathrm{NN}}$-dimensional variables can be expressed as follows:
            \begin{equation}
                \boldsymbol{h}\left(\boldsymbol{z}_{\mathrm{NN}}\right) = \boldsymbol{W}\boldsymbol{z}_{\mathrm{NN}} + \boldsymbol{b} \enspace,
                \label{eq:linear}
            \end{equation}
            where $\boldsymbol{z}_{\mathrm{NN}} \in \mathbb{R}^{N_{\mathrm{NN}}}$ is the input vector, $\boldsymbol{W} \in \mathbb{R}^{M_{\mathrm{NN}} \times N_{\mathrm{NN}}}$ and $\boldsymbol{b} \in \mathbb{R}^{M_{\mathrm{NN}}}$ represent the weight matrix and bias vector of the fully connected layer, respectively. 
            The activation function is denoted by $g$, where functions such as the hyperbolic tangent function ($\tanh$) and the rectified linear function (ReLU) are commonly used. Each variable is converted using the activation function as follows:
            \begin{equation}
                \boldsymbol{g}\left(\boldsymbol{z}\right) = \left( g(z_{1}), g(z_{2}), \cdots, g(z_{M})\right)^{\mathsf{T}}
                \enspace.
                \label{eq:activation}
            \end{equation}
            
            An FNN consists of multiple linear transform functions and activation functions. An FNN with $L$ hidden layers is defined as follows:
            \begin{equation}
                \boldsymbol{y}_{\mathrm{NN}} = \boldsymbol{h}_{L+1} \circ \boldsymbol{g}_{L} \circ \boldsymbol{h}_{L} \circ \cdots \circ \boldsymbol{g}_{1} \circ \boldsymbol{h}_{1} \left(\boldsymbol{x}_{\mathrm{NN}}\right) \enspace,
                \label{eq:mlp}
            \end{equation}
            where $\circ$ denotes the composition operator, $\boldsymbol{y}_{\mathrm{NN}}$ is the output vector, and $\boldsymbol{x}_{\mathrm{NN}}$ is the input vector. It is important to note that there is no activation function applied in the final layer. 
            For simplicity, the FNN can be represented as:
            \begin{equation}
                \boldsymbol{y}_{\mathrm{NN}} = \boldsymbol{\phi}_{\boldsymbol{\theta}}\left(\boldsymbol{x}_{\mathrm{NN}}\right) 
                \enspace,
                \label{eq:mlp_phi}
            \end{equation}
            where $\boldsymbol{\theta}$ is the parameter vector that includes all weight matrices $\boldsymbol{W}$ and bias vectors $\boldsymbol{b}$ for each layer.
    
    \section{Ensemble learning for ship maneuvering motion}\label{sec:learning}
        In this section, we describe the Ensemble learning method using ANNs for the probabilistic prediction of ship's maneuvering motions. Ensemble learning is a machine learning algorithm that constructs a set of models, called ensemble members, and computes a new prediction based on the individual predictions of each member \cite{Dietterich2000,Sagi2018}. In the context of ANN, the ensemble method is considered an approximation of Bayesian inference, capable of capturing epistemic uncertainty caused by limited data. It is a scalable approach that handles large datasets \cite{Lakshminarayanan2017,rahaman2021}.
        
        In this method, we assume that only multiple trajectory datasets composed of the kinematic state variables $\boldsymbol{x}$ and the input variables $\left(\boldsymbol{u}, \boldsymbol{w}_{\mathrm{T}}\right)$ are provided, without any observations of accelerations or forces. Here, the term "trajectory data" refers to sequence data of state variables defined as a function of time. Through maximum likelihood estimation, we obtain plausible maneuvering model $\boldsymbol{F}$ under a given dataset. The maneuvering model $\boldsymbol{F}$ is represented using FNN, and the likelihood is defined by comparing the observed and simulated kinematic state. This approach was introduced in previous works \cite{Wakita2022,Wakita2024}, where its effectiveness was demonstrated using noisy data observed from a model ship. However, the proposed method differs in the following aspects:
        \begin{itemize}
            \item To improve the fitting accuracy of the maneuvering model, an ANN is introduced for the estimation of the initial state.
            \item To increase the diversity among ensemble members, regularization terms in the loss function have been excluded, and early stopping using validation data has been eliminated.
        \end{itemize}
        The following sections provide detailed explanations of the structure of the maneuvering model in \Cref{subsec:FNN_structure}, the loss function of maneuvering model in \Cref{subsec:loss}, and the optimization method in \Cref{subsec:opt}.
    
        \subsection{Maneuvering model structure}\label{subsec:FNN_structure}
            The maneuvering model $\boldsymbol{F}$ is modeled by a single FNN $\boldsymbol{\phi}_{\boldsymbol{\theta}}: \mathbb{R}^{8} \rightarrow \mathbb{R}^{3}$ parameterized by $\boldsymbol{\theta}$. However, certain enhancements have been applied to the pre- and post-processing of the FNN, as detailed below.
    
            \begin{itemize}
                \item Periodic variables are removed from the input variables $\boldsymbol{\nu}, \boldsymbol{u}, \boldsymbol{w}_{\mathrm{A}}$ of the maneuvering model $\boldsymbol{F}$. Specifically, the apparent wind speed and direction $\left(U_{A}, \gamma_{A}\right)$ are transformed into the apparent wind velocity vector $\boldsymbol{w}_{\mathrm{A}}^{\prime}$. As a result, the input variables are represented as follows:
                \begin{equation}
                    \boldsymbol{x}_{\boldsymbol{F}} = \left(\boldsymbol{\nu}^{\mathsf{T}}, \boldsymbol{u}^{\mathsf{T}}, U_{A}\cos\gamma_{A}, U_{A}\sin\gamma_{A}\right)^{\mathsf{T}} \in \mathbb{R}^{8} \enspace.
                    \label{eq:input_dynamics}
                \end{equation}
                %
                \item The input variables are standardized using the training data. Specifically, let $\bar{\boldsymbol{x}}_{\boldsymbol{F}}$ represent the standardized variables, and let $x_{\boldsymbol{F}}$ and $\bar{x}_{\boldsymbol{F}}$ denote a particular component of $\boldsymbol{x}_{\boldsymbol{F}}$ and $\bar{\boldsymbol{x}}_{\boldsymbol{F}}$, respectively. The standardized variables are expressed as follows:
                \begin{equation}
                    \bar{x}_{\boldsymbol{F},i} = \frac{x_{\boldsymbol{F},i}-\mu^{(\text{train})}_{x_{\boldsymbol{F},i}}}{\sigma^{(\text{train})}_{x_{\boldsymbol{F},i}}} \enspace,
                    \label{eq:std_input_dynamics}
                \end{equation}
                where $\mu^{(\text{train})}_{x_{\boldsymbol{F}}}$ and $\sigma^{(\text{train})}_{x_{\boldsymbol{F}}}$ represent the mean and standard deviation of the variable $x_{\boldsymbol{F}}$ in the training data, respectively. As a result, $\bar{\boldsymbol{x}}_{\boldsymbol{F}}$ serves as the input to the FNN.
                %
                \item The FNN output is linearly transformed to match the mean and standard deviation of the accelerations computed from the training dataset. Specifically, first, the mean and standard deviation of the FNN output are calculated using input samples from a standard normal distribution, and the FNN output is standardized using that average and standard deviation. Let $\boldsymbol{y}_{\boldsymbol{F}} \in \mathbb{R}^{3}$ denote the output variables of the FNN, and let $y_{\boldsymbol{F}}$ represent one component of $\boldsymbol{y}_{\boldsymbol{F}}$. The standardized output is expressed as follows:
                \begin{equation}
                    \bar{y}_{\boldsymbol{F}} = \frac{y_{\boldsymbol{F}}-\mu^{(\text{normal})}_{y_{\boldsymbol{F}}}}{\sigma^{(\text{normal})}_{y_{\boldsymbol{F}}}} \enspace,
                    \label{eq:std_output_dynamics}
                \end{equation}
                where $\mu^{(\text{normal})}_{y_{\boldsymbol{F}}}$ and $\sigma^{(\text{normal})}_{y_{\boldsymbol{F}}}$ represent the mean and standard deviation of the FNN output when samples from a standard normal distribution are used as input. In this study, $10,000$ samples were used. 
                Next, to reduce the scale differences among output elements, inverse standardization is performed based on the mean and standard deviation of the ship's accelerations calculated from the training data using numerical time derivatives. Denoting a component of model output $\dot{\boldsymbol{\nu}}$ by $\dot{\nu}$, it is calculated as follows:
                \begin{equation}
                    \dot{\nu} = \bar{y}_{\boldsymbol{F}} \cdot \sigma^{(\text{train})}_{\dot{\nu}} + \mu^{(\text{train})}_{\dot{\nu}} \enspace,
                    \label{eq:invstd_output_dynamics}
                \end{equation}
                where $\mu^{(\text{train})}_{\dot{\nu}}$ and $\sigma^{(\text{train})}_{\dot{\nu}}$ represent the mean and standard deviation of $\dot{\nu}$.
            \end{itemize}
        
            In the following, the maneuvering model $\boldsymbol{F}$ expressed by the FNN is denoted as follows:
            \begin{equation}
                \dot{\boldsymbol{\nu}} = \boldsymbol{F}_{\boldsymbol{\theta}}\left(\boldsymbol{\nu}, \boldsymbol{u}, \boldsymbol{w}_{\mathrm{A}}\right) 
                \enspace.
                \label{eq:FNN_dynamics}
            \end{equation}
            Additionally, \Cref{eq:wT2wA_dynamics} and \Cref{eq:concat_kinamatics_and_dynamics}, which are derived from \Cref{eq:FNN_dynamics}, are respectively expressed as follows: 
            \begin{subequations}\label{eq:FNN_kinamatics_and_dynamics}
                \begin{align}
                \dot{\boldsymbol{\nu}} &= \boldsymbol{F}^{\prime}_{\boldsymbol{\theta}}\left(\boldsymbol{x}, \boldsymbol{u}, \boldsymbol{w}_{\mathrm{T}}\right) \enspace, \label{eq:FNN_wT2wA_dynamics} \\
                \dot{\boldsymbol{x}} &= \boldsymbol{f}_{\boldsymbol{\theta}}\left(\boldsymbol{x}, \boldsymbol{u}, \boldsymbol{w}_{\mathrm{T}}\right) \enspace. \label{eq:FNN_concat_kinamatics_and_dynamics} 
                \end{align}
            \end{subequations}
            
        \subsection{Loss function for learning a model}\label{subsec:loss}
            FNN parameters $\boldsymbol{\theta}$ are determined by maximum likelihood estimation on multiple trajectory datasets, under the assumption that the observed values of the ship's kinematic state $\boldsymbol{x}$ are samples with additive Gaussian noise applied to the true values. In the proposed method, compared to the previous study \cite{Wakita2022}, an ANN is introduced to estimate the initial values of the kinematic state $\boldsymbol{x}$. \Cref{subsubsec:likelihood_a_traj} describe the likelihood function for a single trajectory, \Cref{subsubsec:initial_state_estimation} describe the ANN model for initial state estimation, and \Cref{subsubsec:likelihood_multi_traj} describe the likelihood function for multiple trajectory datasets, as well as the final loss function used for optimization.
    
            \subsubsection{Likelihood function given a observed trajectory}\label{subsubsec:likelihood_a_traj}
                First, the likelihood function for a single trajectory dataset is described. Let the observation of the kinematic state $\boldsymbol{x}$ be denoted as $\boldsymbol{y}$. In this method, it is assumed that time-series data of observations $\boldsymbol{y}$ and the input variables $\boldsymbol{u}, \boldsymbol{w}_{\mathrm{T}}$ are given. Each of these variables is observed at $K$ discrete time points $\left\{t_{k}\right\}^{K-1}_{k=0}$, and their sequences are defined as follows.
                \begin{equation}
                    \left\{
                    \begin{aligned}
                    \boldsymbol{Y} &= \left(\boldsymbol{y}^{\mathsf{T}}\left(t_{0}\right), \boldsymbol{y}^{\mathsf{T}}\left(t_{1}\right), \cdots, \boldsymbol{y}^{\mathsf{T}}\left(t_{K-1}\right)\right)^{\mathsf{T}} \in \mathbb{R}^{6K} \\
                    \boldsymbol{U} &= \left(\boldsymbol{u}^{\mathsf{T}}\left(t_{0}\right), \boldsymbol{u}^{\mathsf{T}}\left(t_{1}\right), \cdots, \boldsymbol{u}^{\mathsf{T}}\left(t_{K-1}\right)\right)^{\mathsf{T}}\in \mathbb{R}^{3K} \\
                    \boldsymbol{W}_{\mathrm{T}} &= \left(\boldsymbol{w}_{\mathrm{T}}^{\mathsf{T}}\left(t_{0}\right), \boldsymbol{w}_{\mathrm{T}}^{\mathsf{T}}\left(t_{1}\right), \cdots, \boldsymbol{w}_{\mathrm{T}}^{\mathsf{T}}\left(t_{K-1}\right)\right)^{\mathsf{T}}\in \mathbb{R}^{2K}
                    \enspace.
                    \end{aligned}
                    \right.
                    \label{eq:sequence_vectors}
                \end{equation}
                Here, the time step does not necessarily have to be constant.
    
                The observation $\boldsymbol{y}$ is assumed to be a sample with Gaussian noise added to the true value. That is, the observation $\boldsymbol{y}$ is defined as follows:
                \begin{equation}
                    \boldsymbol{y} = \boldsymbol{x} + \boldsymbol{\epsilon} \enspace,
                    \label{eq:obs_eq}
                \end{equation}
                where, $\boldsymbol{\epsilon} \sim \mathcal{N}\left(\boldsymbol{0}, \boldsymbol{\Sigma}_{\boldsymbol{\epsilon}}\right)$ represents noise following a zero-mean multivariate Gaussian distribution, with a known covariance matrix $\boldsymbol{\Sigma}_{\boldsymbol{\epsilon}} \in \mathbb{R}^{6 \times 6}$.
                The Gaussian noise $\boldsymbol{\epsilon}$ is assumed to be temporally and spatially independent. Specifically, let the standard deviations of each component of $\boldsymbol{x}$ be denoted as $\sigma_{x_0}, \sigma_{y_0}, \sigma_{\psi}, \sigma_{u}, \sigma_{v_{\mathrm{m}}}, \sigma_{r}$ respectively, the covariance matrix is defined as $\boldsymbol{\Sigma}_{\boldsymbol{\epsilon}}=\mathrm{diag}\left(\sigma_{x_0}^{2}, \sigma_{y_0}^{2}, \sigma_{\psi}^{2}, \sigma_{u}^{2}, \sigma_{v_{\mathrm{m}}}^{2}, \sigma_{r}^{2}\right)^{\mathsf{T}}$.
                The standard deviations of the velocity components $\sigma_{u}, \sigma_{v_{\mathrm{m}}}, \sigma_{r}$ are determined based on the catalogue specifications of the observation equipment. As a result of trial and error, in this paper, we set $\sigma_{x_0}=\sigma_{y_0}=\sigma_{\psi}=\infty$ in training and ignored the error in the pose components.
    
                The kinematic state $\boldsymbol{x}$ can be estimated using the maneuvering model. Specifically, the estimation at a given time $t$ is determined as follows.
                \begin{equation}
                    \boldsymbol{x}_{\boldsymbol{\theta}, \boldsymbol{x}_{0}}\left(t\right) = \boldsymbol{x}_{0} + \int^{t}_{\tau=t_{0}}\boldsymbol{f}_{\boldsymbol{\theta}}\left(\boldsymbol{x}_{\boldsymbol{\theta}, \boldsymbol{x}_{0}}\left(\tau\right), \boldsymbol{u}\left(\tau\right),\boldsymbol{w}_{\mathrm{T}}\left(\tau\right)\right) \mathrm{d}\tau \enspace.
                    \label{eq:estimated_x}
                \end{equation}
                Here, $\boldsymbol{x}_{0}$ is a parameter that represents the initial state, and the subscripts in $\boldsymbol{x}_{\boldsymbol{\theta}, \boldsymbol{x}_{0}}$ indicate that they depend on $\boldsymbol{\theta}$ and $\boldsymbol{x}_{0}$. In addition, the integral in \Cref{eq:estimated_x} is numerically solved using the Euler method.
                
                
                Thus, given the parameters $\left(\boldsymbol{\theta}, \boldsymbol{x}_{0}\right)$, the likelihood function for the observation sequence $\boldsymbol{Y}$ is defined as follows.
                \begin{equation}
                    p\left( \boldsymbol{Y} \mid \boldsymbol{\theta}, \boldsymbol{x}_{0} \right) = \prod_{k=0}^{K-1}\mathcal{N}\left( \boldsymbol{y}\left(t_{k}\right) \mid \boldsymbol{x}_{\boldsymbol{\theta}, \boldsymbol{x}_{0}}\left(t_{k}\right), \boldsymbol{\Sigma}_{\boldsymbol{\epsilon}} \right) \enspace.
                    \label{eq:likelihood_a_traj}
                \end{equation}
                
            \subsubsection{Initial state estimation}\label{subsubsec:initial_state_estimation}
                To obtain the likelihood function $p\left( \boldsymbol{Y} \mid \boldsymbol{\theta}, \boldsymbol{x}_{0} \right)$, it is necessary to have the parameter $\boldsymbol{x}_{0}$. As will be discussed later, when considering the likelihood function for multiple trajectories, the number of parameters varies depending on the number of trajectories. To address this, in the proposed method, the initial values $\boldsymbol{x}_{0}$ are expressed using a FNN that takes the trajectory data as input. This enables a single FNN to represent the estimated initial state for all trajectories. 
                However, it is not always necessary to use the entire observation sequence $\boldsymbol{Y}$ as input. Therefore, the number of time steps used is set to $K_{\mathrm{init}} < K$, and the observation sequence for the initial state estimation is denoted as follows:
                \begin{equation}
                    \boldsymbol{Y}_{\mathrm{init}} = \left(\boldsymbol{y}^{\mathsf{T}}\left(t_{0}\right), \boldsymbol{y}^{\mathsf{T}}\left(t_{1}\right), \cdots, \boldsymbol{y}^{\mathsf{T}}\left(t_{K_{\mathrm{init}}-1}\right)\right)^{\mathsf{T}} \in \mathbb{R}^{6K_{\mathrm{init}}} \enspace.
                    \label{eq:Y_init}
                \end{equation}
                Thus, the initial state $\boldsymbol{x}_{0}$ are determined as follows:
                \begin{equation}
                    \boldsymbol{x}_{0} = \boldsymbol{y}\left(t_{0}\right) + \boldsymbol{\phi}^{\prime}_{\boldsymbol{\theta}^{\prime}}\left(\boldsymbol{Y}_{\mathrm{init}}\right) \enspace,
                    \label{eq:estimated_init_state}
                \end{equation}
                where $\boldsymbol{\phi}^{\prime}_{\boldsymbol{\theta}^{\prime}} : \mathbb{R}^{6 \times K} \rightarrow \mathbb{R}^{6}$ is an FNN parameterized by $\boldsymbol{\theta}^{\prime}$. That is, the kinematic state $\boldsymbol{x}$ is estimated as follows:
                \begin{equation}
                    \begin{aligned}
                        \boldsymbol{x}_{\boldsymbol{\theta}, \boldsymbol{\theta}^{\prime}}\left(t\right) = & \boldsymbol{y}\left(t_{0}\right) + \boldsymbol{\phi}^{\prime}_{\boldsymbol{\theta}^{\prime}}\left(\boldsymbol{Y}_{\mathrm{init}}\right) \\
                        & + \int^{t}_{\tau=t_{0}}\boldsymbol{f}_{\boldsymbol{\theta}}\left(\boldsymbol{x}_{\boldsymbol{\theta}, \boldsymbol{\theta}^{\prime}}\left(\tau\right), \boldsymbol{u}\left(\tau\right), \boldsymbol{w}_{\mathrm{T}}\left(\tau\right)\right) \mathrm{d}\tau \enspace. \\
                    \end{aligned}
                    \label{eq:estimated_x_2}
                \end{equation}
                Therefore, given the parameters $\left(\boldsymbol{\theta}, \boldsymbol{\theta}^{\prime}\right)$, the likelihood function is redefined as follows:
                \begin{equation}
                    p\left( \boldsymbol{Y} \mid \boldsymbol{\theta}, \boldsymbol{\theta}^{\prime} \right) = \prod_{k=0}^{K-1}\mathcal{N}\left( \boldsymbol{y}\left(t_{k}\right) \mid \boldsymbol{x}_{\boldsymbol{\theta}, \boldsymbol{\theta}^{\prime}}\left(t_{k}\right), \boldsymbol{\Sigma}_{\boldsymbol{\epsilon}} \right) \enspace.
                \end{equation}
            
            \subsubsection{Likelihood function given observed trajectories}\label{subsubsec:likelihood_multi_traj}
                The loss function is defined using a likelihood function for multiple trajectories.
                Let us assume that we are given $N$ sequences of observations $\boldsymbol{Y}$, input variables $\boldsymbol{U}$, and external force variables $\boldsymbol{W}_{\mathrm{T}}$. The set of these data is defined as follows:
                \begin{equation}
                    \mathcal{D} = \left\{ \left( \boldsymbol{Y}_{n}, \boldsymbol{U}_{n}, \boldsymbol{W}_{\mathrm{T},n}\right)\right\}_{n=1,\cdots,N} \enspace.
                    \label{eq:dataset}
                \end{equation}
                Note that the $k$-th time point in $\boldsymbol{Y}_{n}$, $\boldsymbol{U}_{n}$ and $\boldsymbol{W}_{\mathrm{T},n}$ is denoted by $t_{n,k}$ in the following.
                The likelihood function of the dataset $\mathcal{D}$ is then defined as follows:
                \begin{equation}
                    p\left( \mathcal{D} \mid \boldsymbol{\theta}, \boldsymbol{\theta}^{\prime} \right) = \prod_{n=1}^{N}\prod_{k=0}^{K-1}\mathcal{N}\left( \boldsymbol{y}\left(t_{n,k}\right) \mid \boldsymbol{x}_{\boldsymbol{\theta}, \boldsymbol{\theta}^{\prime}}\left(t_{n,k}\right), \boldsymbol{\Sigma}_{\boldsymbol{\epsilon}} \right) \enspace.
                    \label{eq:likelihood_multi_traj}
                \end{equation}
                Therefore, the negative log-likelihood function for multiple trajectory data is defined as follows:
                \begin{equation}
                    \begin{aligned}
                        \mathcal{L} & \left( \boldsymbol{\theta}, \boldsymbol{\theta}^{\prime}; \mathcal{D} \right)  \\
                        & = - \log p\left( \mathcal{D} \mid \boldsymbol{\theta}, \boldsymbol{\theta}^{\prime} \right) \\
                        & = - \sum_{n=1}^{N}\sum_{k=0}^{K-1}\log\mathcal{N}\left( \boldsymbol{y}\left(t_{n,k}\right) \mid \boldsymbol{x}_{\boldsymbol{\theta}, \boldsymbol{\theta}^{\prime}}\left(t_{n,k}\right), \boldsymbol{\Sigma}_{\boldsymbol{\epsilon}} \right) +(\mathrm{const.}) \\
                        & = \sum_{n=1}^{N}\sum_{k=0}^{K-1} 
                        \left\|\boldsymbol{y}\left(t_{n,k}\right)-\boldsymbol{x}_{\boldsymbol{\theta}, \boldsymbol{\theta}^{\prime}}\left(t_{n,k}\right)\right\|_{\boldsymbol{\Sigma}_{\boldsymbol{\epsilon}}^{-1}}^{2}
                        +(\mathrm{const.}) \enspace.
                    \end{aligned}
                    \label{eq:loss_func}
                \end{equation}
                Here, constant terms that do not depend on the parameters $\left( \boldsymbol{\theta}, \boldsymbol{\theta}^{\prime} \right)$ have been omitted.
                This negative log-likelihood function is used for loss function.
                
        \subsection{Optimization methods}\label{subsec:opt}
            In this method, multiple maneuvering models are trained by finding the FNN parameters $\left( \boldsymbol{\theta}, \boldsymbol{\theta}^{\prime} \right)$ that minimize the loss function given in \Cref{eq:loss_func}. The minimization is achieved using the Adam algorithm \cite{kingma2014}, a gradient descent-based method, in combination with mini-batch learning, where the dataset is divided into smaller batches for gradient computation. 

            
            Techniques such as Bagging and Boosting \cite{Livieris2021}, and Data Augmentation \cite{wen2021} have been proposed to enhance model diversity. When using neural networks, it is expected that many local optima exist. Studies have demonstrated that the random initialization of neural network parameters and random shuffling of training data can significantly increase diversity among neural network models \cite{lee2015, Lakshminarayanan2017}. Therefore, although the same dataset $\mathcal{D}$ is used for training all maneuvering models, the order of mini-batch data is randomized during training, and the initial values of FNN parameters are determined randomly. In the implementation, the default weight initialization in PyTorch was utilized.
    
            Hereafter, the set of $M$ optimal parameter vectors is denoted as $\Theta \equiv \left\{ \boldsymbol{\theta}_{m}^{\star} \right\}_{m=1}^{M}$, and the set of $M$ optimal maneuvering models is denoted as $\Phi \equiv \left\{\boldsymbol{f}_{\boldsymbol{\theta}} \mid \boldsymbol{\theta} \in \Theta\right\}$.
    
    \section{Maneuvering motion prediction with trained maneuvering models}\label{sec:prediction}
        This section describes the prediction method of maneuvering motion using the set of models $\Phi$ obtained from proposed ensemble learning. In the proposed method, a maneuvering model is selected with uniform probability from the model set $\Phi$, and the state variables are propagated using particle method based on that maneuvering model, enabling for probabilistic predictions of maneuvering motion.
    
        Let the initial state variables $\boldsymbol{x}\left(t_{0}\right)$ and the input variable sequence $\left(\boldsymbol{u}\left(t_{0}\right),\boldsymbol{w}_{\mathrm{T}}\left(t_{0}\right),\boldsymbol{u}\left(t_{1}\right),\boldsymbol{w}_{\mathrm{T}}\left(t_{1}\right), \cdots, \boldsymbol{u}\left(t_{K-1}\right),\boldsymbol{w}_{\mathrm{T}}\left(t_{K-1}\right)\right)$ are given.
        Initially, $P$ particles $\left\{\boldsymbol{x}_{p}\left(t_{0}\right)\right\}_{p=1}^{P}$ are first generated, each starting from the initial state $\boldsymbol{x}\left(t_{0}\right)$. Then, at each time step, a particle $\boldsymbol{x}_{p}\left(t_{k}\right)$ is propagated as follows:
        \begin{equation}
             \boldsymbol{x}_{p}\left(t_{k+1}\right) = \boldsymbol{x}_{p}\left(t_{k}\right) + \int^{t_{k+1}}_{\tau=t_{k}}\boldsymbol{f}_{\boldsymbol{\theta}^{\star}}\left(\boldsymbol{x}_{p}\left(\tau\right), \boldsymbol{u}\left(\tau\right),\boldsymbol{w}_{\mathrm{T}}\left(\tau\right)\right) \mathrm{d}\tau \enspace,
            \label{eq:prediction}
        \end{equation}
        where $\boldsymbol{f}_{\boldsymbol{\theta}^{\star}}$ is the maneuvering model uniformly selected from the model set $\Phi$. The integrals are also solved numerically using the $4$-th order Runge-Kutta method, as in the training.
        
        Chua et al. \cite{Chua2018} proposed two trajectory sampling methods using ensemble models:
        \begin{itemize}
            \item TS$1$ is a method in which the model is resampled at each time step for a particle.
            \item TS$\infty$ is a method in which the model is sampled only once for a particle.
        \end{itemize}
        The manuvering model set $\Phi$ represents a plausible sample set from the function space of the true manuvering model $\boldsymbol{f}$, which is assumed to be time-invariant, based on the training dataset.
        TS$1$ propagates particles using a different maneuvering model at each time step. Although this may appear to violate the time-invariance assumption of the maneuvering model, the little possibility of multiple transitions to the same continuous state in trajectory sampling allows TS$1$ to be considered effectively time-invariant in practical use. Furthermore, TS$1$ is capable of obtaining more plausible trajectory samples than TS$\infty$. However, since TS$1$ has the effect of averaging out predicted states, TS$\infty$ is more effective at capturing epistemic uncertainty when the sample size is limited. Consequently, TS$\infty$ is employed.
    
    \section{Experiments using simulation data of a model ship}\label{sec:sim_exp}
        In this section, to demonstrate the effectiveness of the proposed method, we present the results of numerical experiments on the prediction accuracy of maneuvering motion using numerical simulation data of a model ship. In \Cref{subsec:sim_dataset}, the dataset used in the experiments is described. In \Cref{subsec:comp_init}, we present the changes in fitting accuracy to demonstrate the impact of the FNN on initial state estimation. Then, in \Cref{subsec:sim_pred}, we show the relationship between the prediction accuracy of maneuvering motions and the uncertainty in those predictions. Finally, in \Cref{subsec:ctrl}, we present the results of evaluating the PD gains of heading-keeping PD control using the model obtained in \Cref{subsec:sim_pred}.
        
        \subsection{Dataset}\label{subsec:sim_dataset}
            This section describes the method for generating simulation data of the 3-meter model ship used in the experiment. The maneuvering simulations employed the MMG model, actuator response model, and wind stochastic process model, which are described in \Cref{subsec:mmg_model,subsec:act_model,subsec:wind_model} and were also used in previous research \cite{Wakita2024MBRL}. Here, the MMG model and actuator response model were numerically solved using the 4th-order Runge-Kutta method, while the wind stochastic process model was solved using the Euler-Maruyama method. Additionally, the observed values of kinematic state variables $\boldsymbol{x}$ were artificially polluted with additive Gaussian noise applied to the simulated values, with the noise's standard deviation assumed to be known. The parameters used in the simulation are listed in \Cref{tab:paramsim}.
    
            \begin{table}[t]
                \centering
                \caption{Parameters of numerical simulation.}
                \begin{tabular}{ll} \toprule
                    Item & Value \\ \midrule
                    Time step of numerical simulation & $0.1 \ \mathrm{(s)}$ \\
                    Standard deviation of Gaussian noise in $u$ & $0.01 \ \mathrm{(m/s)}$ \\
                    Standard deviation of Gaussian noise in $v_{\mathrm{m}}$ & $0.01 \ \mathrm{(m/s)}$ \\
                    Standard deviation of Gaussian noise in $r$ & $0.1 \ \mathrm{(deg./s)}$ \\
                    \bottomrule
                \end{tabular}
                \label{tab:paramsim}
            \end{table}
    
            In the maneuvering simulations, zigzag, turning, berthing, and random control maneuvers were conducted. The notation for each trajectory dataset is summarized in \Cref{tab:note_maneuver}. Here, the berthing maneuver is a simulated maneuvering motion using the time series of control inputs made during the port navigation data of full-scale ship described in \Cref{subsec:nav_dataset}. The purpose of using the berthing data is not to estimate the maneuvering model of the full-scale ship but to generate data similar to the state distribution observed during berthing operations.
    
            The random control maneuver refers to a simulated maneuver in which the state command of the actuator is randomly changed at regular intervals. In this study, the port rudder angle $\delta_{\mathrm{P}}$ is determined by sampling from $\mathcal{N}(-80, 30^{2})$, with values clipped to the range $[-105, 35]$. Similarly, the starboard rudder angle $\delta_{\mathrm{S}}$ is sampled from $\mathcal{N}(80, 30^{2})$ and clipped to $[-35, 105]$. The bow thruster speed $n_{\mathrm{BT}}$ is sampled from $\mathcal{N}(0, 15^{2})$ and clipped to the range $[-30, 30]$. These values are updated every $20 \ \mathrm{(s)}$.
    
            \begin{table}[t]
                \centering
                \caption{Notations for maneuvers}
                \begin{tabular}{ll}\toprule
                    Notation & Description\\ \midrule
                    $\mathrm{Z}_{\delta/\psi}$ & A $\delta^{\circ}$-$\psi^{\circ}$ zigzag maneuver. \\
                    $\mathrm{T}_{\delta}$ & A $\delta^{\circ}$ turning maneuver. \\
                    $\mathrm{B}_{i=1, \cdots, 41}$ & A berthing maneuver under wind disturbance. \\
                    $\mathrm{R}_{i=1, \cdots, 10}$ & A random maneuver under wind disturbance. \\ 
                    \bottomrule
                \end{tabular}
                \label{tab:note_maneuver}
            \end{table}
    
\begin{figure*}[t]
                \centering
                \includegraphics[width=\linewidth, trim=0 5 0 0, clip]{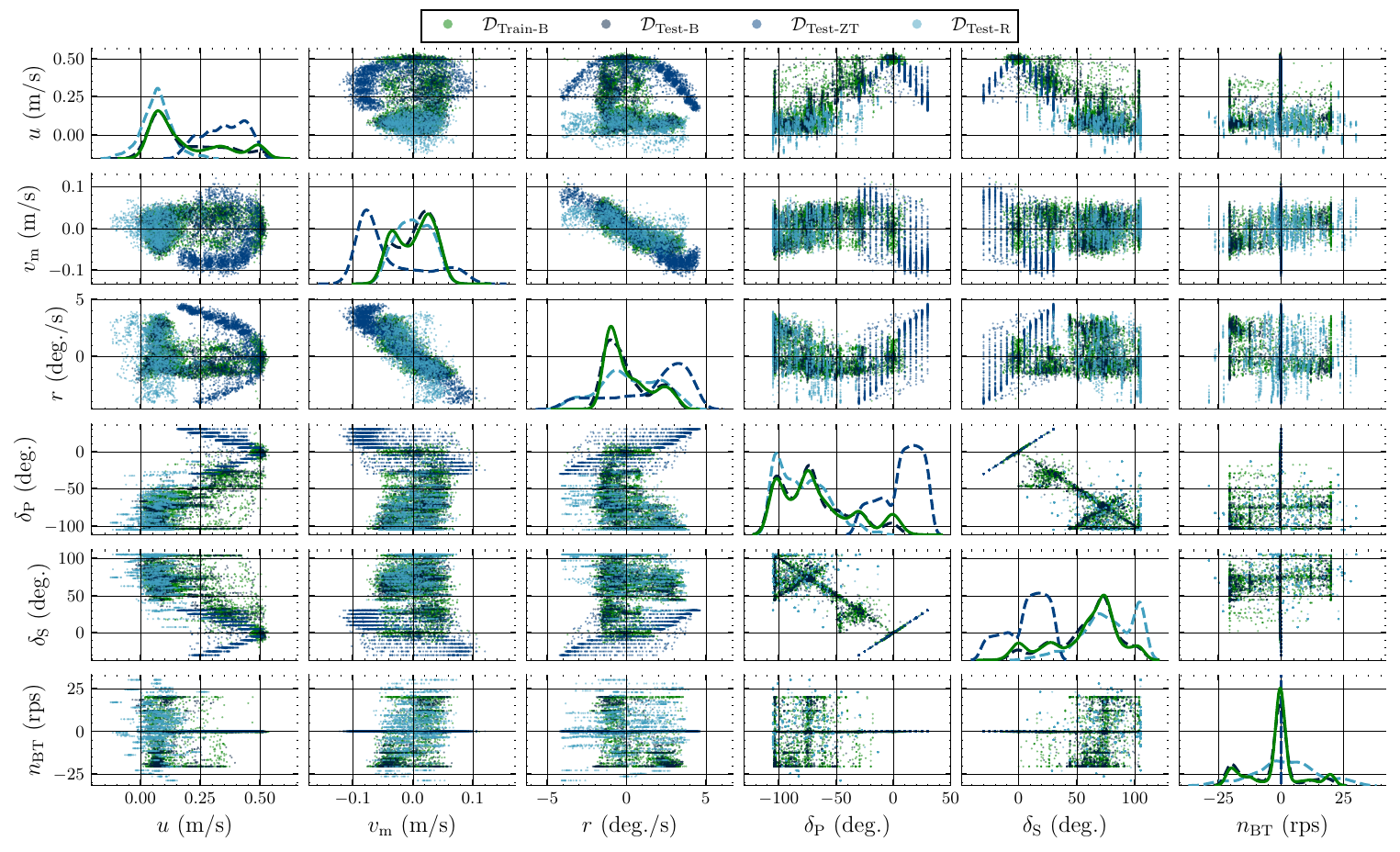}
                \caption{Distribution of ship velocity and actuator state of datasets.}
                \label{fig:pairplot}
            \end{figure*}

            Maneuvers of training and test datasets were selected as shown in \Cref{tab:test_dataset}. The distributions of each dataset are presented in \Cref{fig:pairplot}. Based on these data distributions, the following observations can be made:
            \begin{itemize}
                \item There is no significant difference in the distributions of $\mathcal{D}_{\text{Train-B}}$ and $\mathcal{D}_{\text{Test-B}}$.
                \item The dataset $\mathcal{D}_{\text{Test-ZT}}$ is distributed around regions where the norm of velocity $\boldsymbol{\nu}$ is relatively large. Specifically, it includes large absolute values of sway velocity $v_{\mathrm{m}}$ and yaw velocity $r$, which are not present in $\mathcal{D}_{\text{Train-B}}$.
                \item The dataset $\mathcal{D}_{\text{Test-R}}$ is distributed in regions where the norm of velocity $\boldsymbol{\nu}$ is relatively small. Notably, it includes negative surge velocity, which is absent in $\mathcal{D}_{\text{Train-B}}$.
            \end{itemize}
            The training dataset assumes a scenario where only data collected during berthing operations are available. In such operations, the vessel is maneuvered according to a predetermined pattern to some extent. As shown in \Cref{fig:pairplot}, the distributions of $\mathcal{D}_{\text{Test-ZT}}$ and $\mathcal{D}_{\text{Test-R}}$ are not fully covered by the train dataset. Therefore, there are relatively many state and input variables in the training dataset that are not included in either $\mathcal{D}_{\text{Test-ZT}}$ or $\mathcal{D}_{\text{Test-R}}$.
            
            \begin{table}[t]
                \centering
                \caption{Datasets generated by maneuvering simulation of a model ship}
                \begin{tabular}{lm{40mm}l}\toprule
                    Dataset & Maneuvers & Total Duration \\ \midrule
                    $\mathcal{D}_{\text{Train-B}}$ & $\mathrm{B}_{1}, \mathrm{B}_{2}, \cdots, \mathrm{B}_{27}$  & $4770 \ \mathrm{(s)}$ \\
                    $\mathcal{D}_{\text{Test-B}}$ & $\mathrm{B}_{28}, \mathrm{B}_{29}, \cdots, \mathrm{B}_{41}$ & $2304\ \mathrm{(s)}$ \\
                    $\mathcal{D}_{\text{Test-ZT}}$ & $\mathrm{Z}_{5/5}$, $\mathrm{Z}_{10/10}$, $\mathrm{Z}_{15/15}$, $\mathrm{Z}_{20/20}$, $\mathrm{Z}_{25/25}$, $\mathrm{Z}_{30/30}$, $\mathrm{T}_{5}$, $\mathrm{T}_{10}$, $\mathrm{T}_{15}$, $\mathrm{T}_{20}$, $\mathrm{T}_{25}$, $\mathrm{T}_{30}$ & $2400 \ \mathrm{(s)}$ \\
                    $\mathcal{D}_{\text{Test-R}}$ & $\mathrm{R}_{1}, \mathrm{R}_{2}, \cdots, \mathrm{R}_{10}$ & $2000 \ \mathrm{(s)}$ \\
                    \bottomrule
                \end{tabular}
                \label{tab:test_dataset}
            \end{table}
            
            \begin{table}[t]
                \centering
                \caption{Hyperparameters for training using simulation data of a model ship and navigation data of a full-scale ship.}
                \begin{tabular}{m{40mm}ll}\toprule
                    \multirow{2}{*}{Item} & \multicolumn{2}{c}{Value} \\
                     & Model-scale & Full-scale \\ \midrule
                    Duration of time step: $t_{k+1} - t_{k}$ & $1.0 \ \mathrm{(s)}$ & $7.0 \ \mathrm{(s)}$ \\
                    Number of time steps: $K$ & \multicolumn{2}{l}{$100$} \\
                    Number of time steps for the initial value estimation: $K_{\mathrm{init}}$ &  \multicolumn{2}{l}{$30$} \\
                    Number of FNN hidden layers: $L$ &  \multicolumn{2}{l}{$3$} \\
                    Dimension of FNN hidden layers: $N_{\mathrm{NN}}$ &  \multicolumn{2}{l}{$256$} \\
                    Activation function for FNN hidden layers: $g$ &  \multicolumn{2}{l}{$\tanh$} \\
                    The $u$ component of the standard deviation of $\boldsymbol{\epsilon}$ : $\sigma_{u}$ & $0.01 \ \mathrm{(m/s)}$ & $0.01 \ \mathrm{(m/s)}$ \\
                    The $v_{\mathrm{m}}$ component of the standard deviation of $\boldsymbol{\epsilon}$ : $\sigma_{v_{\mathrm{m}}}$ & $0.01 \ \mathrm{(m/s)}$ & $0.01 \ \mathrm{(m/s)}$ \\
                    The $r$ component of the standard deviation of $\boldsymbol{\epsilon}$ : $\sigma_{r}$ & $0.1 \ \mathrm{(deg./s)}$ & $0.01 \ \mathrm{(deg./s)}$ \\
                    Learning rate &  \multicolumn{2}{l}{$1.0\times10^{-4}$} \\
                    Batch size & \multicolumn{2}{l}{$32$} \\
                    Number of Epochs & $20,000$ & $100,000$ \\
                    \bottomrule
                \end{tabular}
                \label{tab:hyperparam}
            \end{table}
            
        \subsection{Effects of initial state estimation}\label{subsec:comp_init}
            We demonstrate the change in fitting accuracy to the training data $\mathcal{D}_{\text{Train-B}}$ resulting from the introduction of the FNN $\boldsymbol{\phi}^{\prime}_{\boldsymbol{\theta}^{\prime}}$ for initial state estimation. To assess this, training was conducted both with the proposed method, which incorporates the FNN $\boldsymbol{\phi}^{\prime}_{\boldsymbol{\theta}^{\prime}}$, and with a method that determines the initial state as $\boldsymbol{x}_{0} = \boldsymbol{y}\left(t_{0}\right)$ without using the FNN. To account for the effects of randomness in the training process, such as the initial values of the FNN, the training was repeated 10 times under same conditions. The hyperparameters used for the training are summarized in \Cref{tab:hyperparam}.
    
\begin{figure}[t]
                \centering
                \includegraphics[width=0.95\linewidth, trim=10 10 10 10, clip]{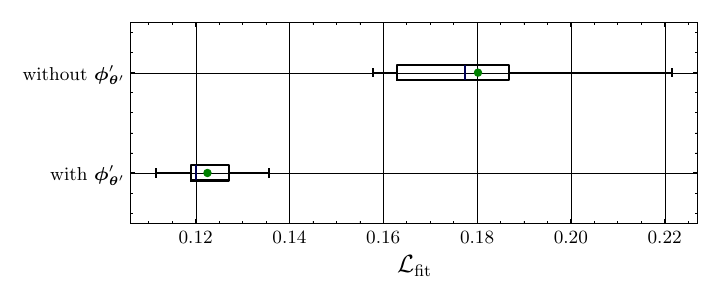}
                \caption{Comparison results of using or not using initial value estimation in fitting accuracy. $\mathcal{L}_{\text{fit}}\left(\mathcal{D}_{\text{Train-B}}\right)$ is a measure of fitting accuracy.}
                \label{fig:boxplot_eval_fitting}
            \end{figure}

            To quantitatively demonstrate the fitting accuracy of the trained models, the following evaluation function was computed for each training result:
            \begin{equation}
                \mathcal{L}_{\text{fit}}\left(\mathcal{D}_{\text{Train-B}} \right) = \frac{1}{NK}\sum_{n=1}^{N}\sum_{k=0}^{K-1}\left\|\boldsymbol{x}\left(t_{n,k}\right)-\boldsymbol{x}_{\boldsymbol{\theta}^{\star}, \boldsymbol{\theta}^{\prime\star}}\left(t_{n,k}\right)\right\|_{\boldsymbol{\Sigma}_{\boldsymbol{\epsilon}}^{-1}}^{2}
                \enspace,
                \label{eq:eval_fitting}
            \end{equation}
            where $\boldsymbol{\Sigma}_{\boldsymbol{\epsilon}}$ is the same value used in training, which means that the pose component is ignored. Note that since $\boldsymbol{x}$ represents the true value of the state variable, $\mathcal{L}_{\text{fit}}$ represents the estimation error from the true value.
            The boxplot of the 10 resulting evaluation function values is shown in \Cref{fig:boxplot_eval_fitting}. From these results, it can be observed that the introduction of the FNN $\boldsymbol{\phi}^{\prime}_{\boldsymbol{\theta}^{\prime}}$ for initial state estimation improves the fitting accuracy of the trained model to the true values.
    
        \subsection{Maneuvering motion prediction and its uncertainty}\label{subsec:sim_pred}
            The prediction accuracy and uncertainty prediction results of the proposed method is demonstrated. In this study, the number of ensembles was set to $M=15$, and the hyperparameters listed in \Cref{tab:hyperparam} were used. The training was conducted five times to account for the randomness inherent in the training process.
            
            For the evaluation of the training results, the test dataset was divided into $K$ steps, resulting in $N$ subsequences of time-series data. Given the initial state variables $\boldsymbol{x}_{n,0}$ from the test dataset, $P$ predicted values of the state variables at each subsequent step, $\left\{\left(\boldsymbol{x}_{n,0,p}, \boldsymbol{x}_{n,1,p}, \cdots, \boldsymbol{x}_{n,K-1,p}\right)\right\}_{p=1, 2, \cdots, P}$, were computed using the proposed method. Here, the number of particles was set to $P=100$, and the number of time steps per trajectory was set to $K=100$.
    
            To quantitatively evaluate the relationship between the prediction accuracy and prediction uncertainty of maneuvering motions, we introduce the mean and covariance of the $P$ predicted particles' velocity components $\boldsymbol{\nu}_{n,k,p}$, defined as follows: 
            \begin{equation}
                \left\{
                \begin{aligned}
                    & \boldsymbol{\mu}_{n,k} = \frac{1}{P}\sum_{p=1}^{P} \boldsymbol{\nu}_{n,k,p} \\
                    & \boldsymbol{\Sigma}_{n,k} = \frac{1}{P} \sum_{p=1}^{P}  \boldsymbol{\nu}_{n,k,p} \boldsymbol{\nu}_{n,k,p}^{\mathsf{T}} - \boldsymbol{\mu}_{n,k} \boldsymbol{\mu}_{n,k}^{\mathsf{T}} \enspace. 
                \end{aligned} 
                \right.
                \label{eq:mean_and_cov}
            \end{equation}
            Next, two metrics are defined to represent the relationship between the set of $P$ predicted particles and the ground truth for the test dataset:
            \begin{subequations} \label{eq:L}
                \begin{align}
                    \mathcal{L}_{\text{Eucl}}\left( \mathcal{D} \right)=&\frac{1}{NK}\sum_{n=1}^{N}\sum_{k=1}^{K} \left\| \boldsymbol{\nu}_{n,k}-\boldsymbol{\mu}_{n,k} \right\|^{2} \label{eq:L_Eucl} \\
                    \mathcal{L}_{\text{Maha}}\left( \mathcal{D} \right)=&\frac{1}{NK}\sum_{n=1}^{N}\sum_{k=1}^{K} \left(\boldsymbol{\nu}_{n,k}-\boldsymbol{\mu}_{n,k}\right)^{\mathsf{T}} \boldsymbol{\Sigma}_{n,k}^{-1} \left(\boldsymbol{\nu}_{n,k}-\boldsymbol{\mu}_{n,k}\right) \enspace.
                \label{eq:L_Maha}
                \end{align}
            \end{subequations}
            Here, $\mathcal{L}_{\text{Eucl}}$ represents the mean squared Euclidean distance between the mean of the predicted particles and the ground truth, while $\mathcal{L}_{\text{Maha}}$ represents the mean squared Mahalanobis distance between the set of predicted particles and the ground truth. The former indicates the magnitude of bias between the predicted particles and the ground truth, while the latter quantifies the bias magnitude normalized by the variance of the prediction set, represented by the covariance matrix. In other words, $\mathcal{L}_{\text{Eucl}}$ serves as a measure of the likelihood that the ground truth belongs to the predicted set.

            \begin{figure}[t]
                \begin{minipage}[c]{\linewidth}
                    \centering
                    \includegraphics[width=\hsize, trim=10 10 10 10, clip]{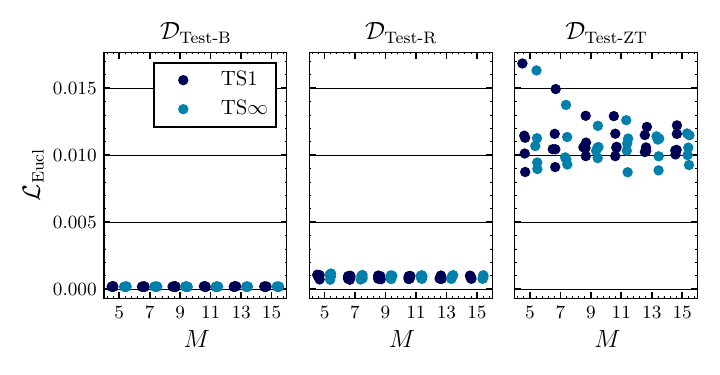}
                    \subcaption{Mean squared Euclidean distance between the mean value of the predicted particles and the true value.}
                    \label{fig:compts_MSE}
                \end{minipage}
                \begin{minipage}[c]{\linewidth}
                    \centering
                    \includegraphics[width=\hsize, trim=10 10 10 10, clip]{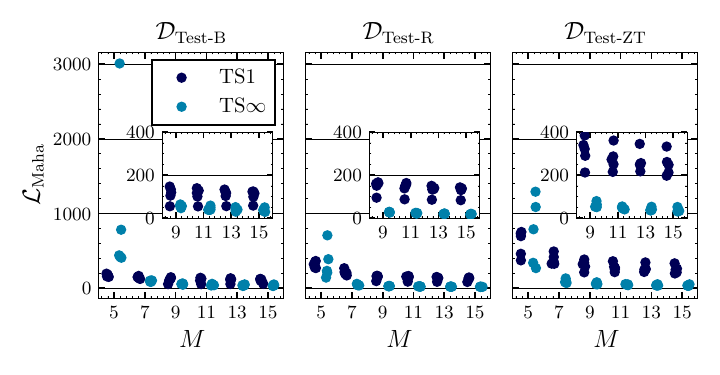}
                    \subcaption{Mean squared Mahalanobis distance between the set of the predicted particles and the true value}
                    \label{fig:compts_NLL}
                \end{minipage}
                \caption{Comparison of prediction accuracy and uncertainty estimation accuracy at $K=100$ for trajectory sampling method comparison.}
                \label{fig:compts}
            \end{figure}

            \begin{figure}[t]
                \centering
                \includegraphics[width=\linewidth, trim=10 10 10 10, clip]{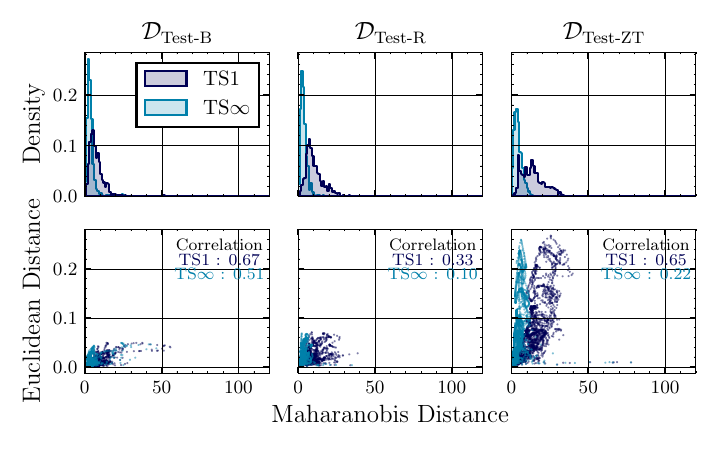}
                \caption{Distribution of Mahalanobis distance and Euclidean distance at $K=100$, $M=15$.}
                \label{fig:compts_dist}
            \end{figure}

            \begin{figure*}[t]
                \centering
                \begin{minipage}[b]{0.32\linewidth}
                    \begin{minipage}[b]{\linewidth}
                        \centering
                        \includegraphics[width=0.95\hsize, trim=10 10 10 10, clip]{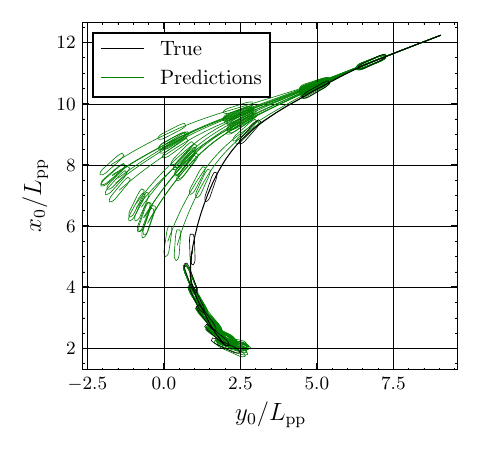}
                        \subcaption{Prediction of ship pose using TS$\infty$.}
                        \label{fig:TSinfty_nav_kanda20220509in_607_x0y0}
                    \end{minipage}
                    \begin{minipage}[b]{\linewidth}
                        \centering
                        \includegraphics[width=0.95\hsize, trim=10 10 10 10, clip]{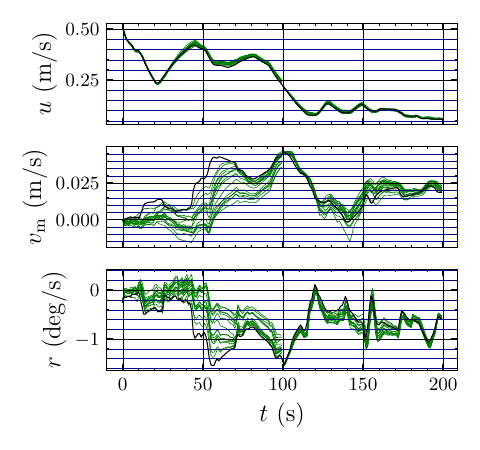}
                        \subcaption{Prediction of ship velocity using TS$\infty$.}
                        \label{fig:TSinfty_nav_kanda20220509in_607_nu}
                    \end{minipage}
                \end{minipage}
                \begin{minipage}[b]{0.32\linewidth}
                    \begin{minipage}[b]{\linewidth}
                        \centering
                        \includegraphics[width=0.95\hsize, trim=10 10 10 10, clip]{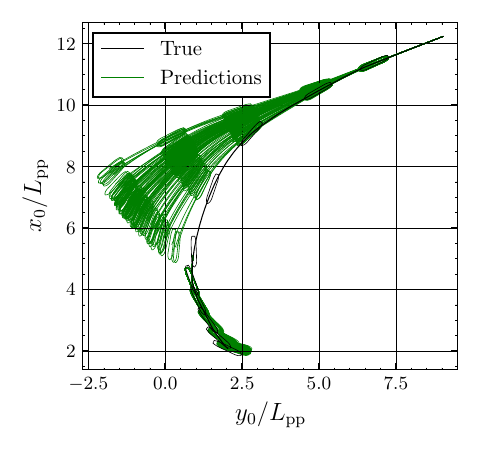}
                        \subcaption{Prediction of ship pose using TS$1$.}
                        \label{fig:TS1_nav_kanda20220509in_607_x0y0}
                    \end{minipage}
                    \begin{minipage}[b]{\linewidth}
                        \centering
                        \includegraphics[width=0.95\hsize, trim=10 10 10 10, clip]{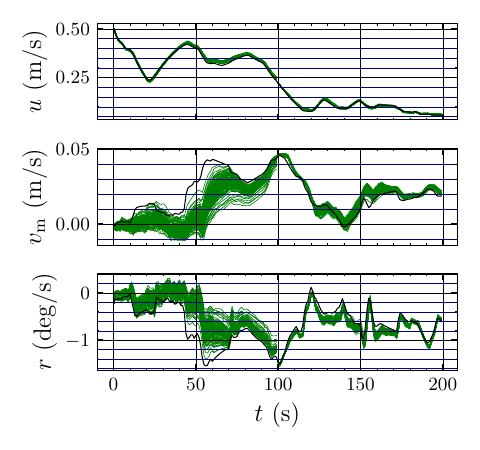}
                        \subcaption{Prediction of ship velocity using TS$1$.}
                        \label{fig:TS1_nav_kanda20220509in_607_nu}
                    \end{minipage}
                \end{minipage}
                \begin{minipage}[b]{0.32\linewidth}
                    \begin{minipage}[b]{\linewidth}
                        \centering
                        \includegraphics[width=0.95\hsize, trim=10 10 10 10, clip]{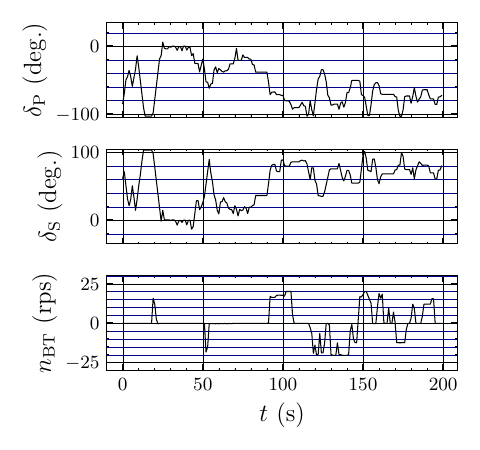}
                        \subcaption{Rudder angles and bow thruster revolution.}
                        \label{fig:TS1_nav_kanda20220509in_607_u}
                    \end{minipage}
                    \begin{minipage}[b]{\linewidth}
                        \centering
                        \includegraphics[width=0.95\hsize, trim=10 10 10 10, clip]{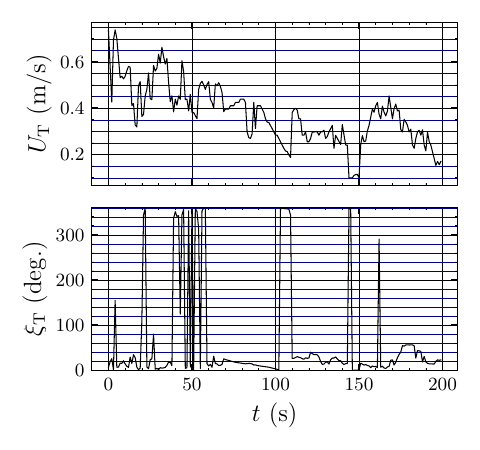}
                        \subcaption{True wind speed and direction.}
                        \label{fig:TS1_nav_kanda20220509in_607_wT}
                    \end{minipage}
                \end{minipage}
                \caption{Prediction of a berthing maneuver. Note that it is initialized using the true value at $t=0.0 \ \mathrm{(s)}$ and $t=100.0 \ \mathrm{(s)}$.}
                \label{fig:nav_kanda20220509in_607}
            \end{figure*}
    
            \begin{figure*}[t]
                \centering
                \begin{minipage}[b]{0.32\linewidth}
                    \begin{minipage}[b]{\linewidth}
                        \centering
                        \includegraphics[width=0.95\hsize, trim=10 10 10 10, clip]{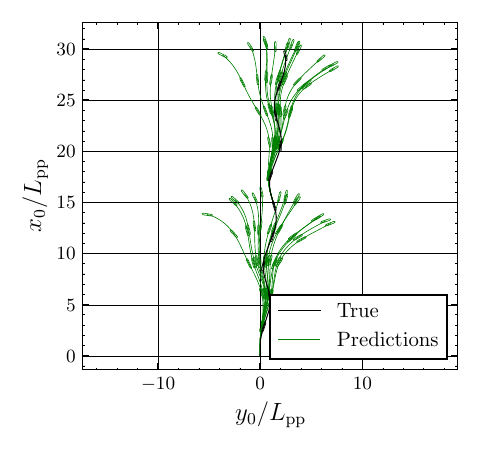}
                        \subcaption{Prediction of ship pose using TS$\infty$.}
                        \label{fig:TSinfty_zigzag_10_10_stab_401_x0y0}
                    \end{minipage}
                    \begin{minipage}[b]{\linewidth}
                        \centering
                        \includegraphics[width=0.95\hsize, trim=10 10 10 10, clip]{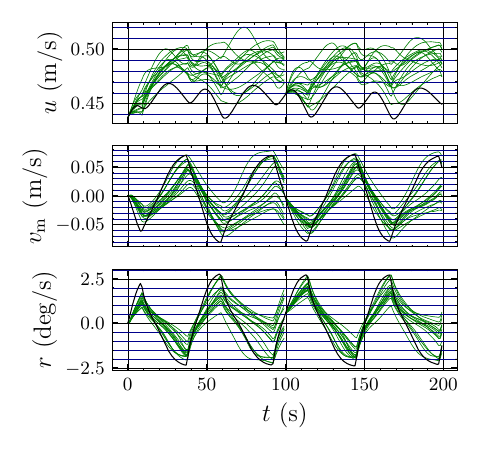}
                        \subcaption{Prediction of ship velocity using TS$\infty$.}
                        \label{fig:TSinfty_zigzag_10_10_stab_401_nu}
                    \end{minipage}
                \end{minipage}
                \begin{minipage}[b]{0.32\linewidth}
                    \begin{minipage}[b]{\linewidth}
                        \centering
                        \includegraphics[width=0.95\hsize, trim=10 10 10 10, clip]{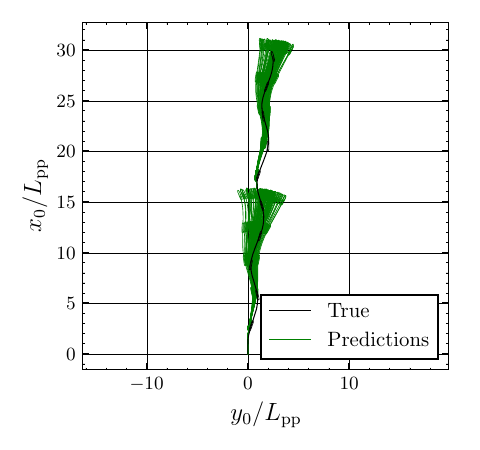}
                        \subcaption{Prediction of ship pose using TS$1$.}
                        \label{fig:TS1_zigzag_10_10_stab_401_x0y0}
                    \end{minipage}
                    \begin{minipage}[b]{\linewidth}
                        \centering
                        \includegraphics[width=0.95\hsize, trim=10 10 10 10, clip]{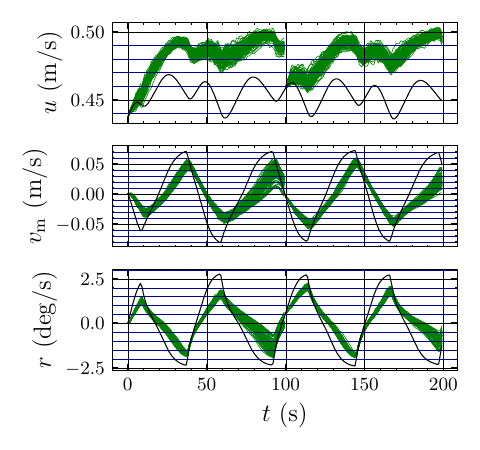}
                        \subcaption{Prediction of ship velocity using TS$1$.}
                        \label{fig:TS1_zigzag_10_10_stab_401_nu}
                    \end{minipage}
                \end{minipage}
                \begin{minipage}[b]{0.32\linewidth}
                    \begin{minipage}[b]{\linewidth}
                        \centering
                        \includegraphics[width=0.95\hsize, trim=10 10 10 10, clip]{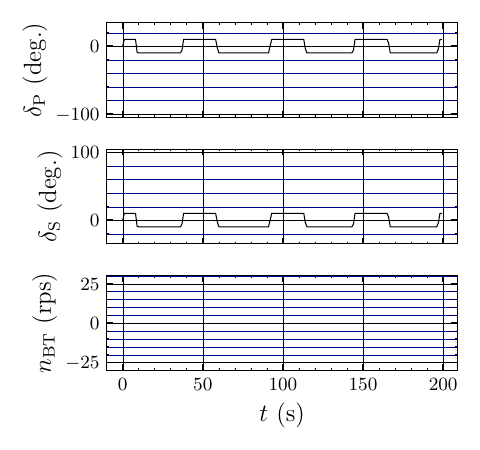}
                        \subcaption{Rudder angles and bow thruster revolution.}
                        \label{fig:TS1_zigzag_10_10_stab_401_u}
                    \end{minipage}
                    \begin{minipage}[b]{\linewidth}
                        \centering
                        \includegraphics[width=0.95\hsize, trim=10 10 10 10, clip]{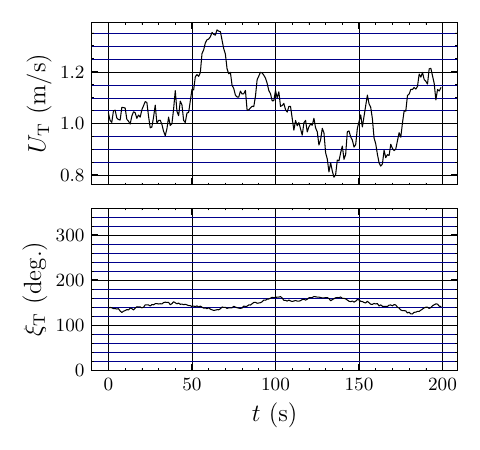}
                        \subcaption{True wind speed and direction.}
                        \label{fig:TS1_zigzag_10_10_stab_401_wT}
                    \end{minipage}
                \end{minipage}
                \caption{Prediction of a $10^{\circ}$-$10^{\circ}$ zigzag maneuver. Note that it is initialized using the true value at $t=0.0 \ \mathrm{(s)}$ and $t=100.0 \ \mathrm{(s)}$.}
                \label{fig:zigzag_10_10_stab_401}
            \end{figure*}

            \Cref{fig:compts} shows the values of $\mathcal{L}_{\text{Eucl}}$ and $\mathcal{L}_{\text{Maha}}$ for the test dataset, TS method and ensemble size $M$. From these results, the following observations can be made:
            \begin{itemize}
                \item Comparison of ensemble sizes: For both methods and all datasets, while $\mathcal{L}_{\text{Eucl}}$ does not change significantly as the ensemble size $M$ increases, $\mathcal{L}_{\text{Maha}}$ decreases and asymptotically approaches a certain value. This suggests that regardless of the method or dataset, increasing the ensemble size $M$ does not reduce the bias of the predicted particle set from the true value, but tends to increase the variance of the predicted particle.
                \item Comparison of TS methods: For all datasets, there is little difference between TS$\infty$ and TS$1$ in terms of $\mathcal{L}_{\text{Eucl}}$. However, when the ensemble size $M$ is sufficiently large, TS$\infty$ exhibits smaller values of $\mathcal{L}_{\text{Maha}}$ than TS$1$. This indicates that when the ensemble size $M$ is large enough, TS$\infty$ does not reduce the bias of the predicted particle set from the true value compared to TS$1$, but tends to increase the variance.
                \item Comparison of datasets: For both methods, when the ensemble size $M$ is sufficiently large, $\mathcal{L}_{\text{Eucl}}\left( \mathcal{D}_{\text{Test-ZT}} \right)$ is more than 10 times larger than $\mathcal{L}_{\text{Eucl}}\left( \mathcal{D}_{\text{Test-B}} \right)$ and $\mathcal{L}_{\text{Eucl}}\left( \mathcal{D}_{\text{Test-R}} \right)$. However, there are no significant differences between datasets in terms of $\mathcal{L}_{\text{Maha}}$. This suggests that as the distribution of the test dataset diverges from that of the training dataset, the bias of the predicted particle set from the true value tends to increase, while the variance of the predicted particle also tends to increase.
            \end{itemize}
    
            Next, \Cref{fig:compts_dist} shows the relationship between the Euclidean distance between the mean of the predicted particles and the true values, and the Mahalanobis distance between the set of predicted particles and the true values, with the ensemble size set to $M=15$. From these results, the following observations can be made:
            \begin{itemize}
                \item For all datasets, the correlation coefficient for TS$\infty$ is smaller compared to TS$1$. This indicates that, compared to TS$1$, TS$\infty$ tends to have smaller variance when the bias of the predicted particle set from the true value is small, and larger variance when the bias is large.
            \end{itemize}

            Finally, prediction results of a Berthing maneuver and a Zigzag maneuver in $\mathcal{D}_{\text{Test-B}}$ and $\mathcal{D}_{\text{Test-ZT}}$ are shown in \Cref{fig:nav_kanda20220509in_607,fig:zigzag_10_10_stab_401}. From these results, the following observations can be drawn:
            \begin{itemize}
                \item The predicted pose varies according to the variation in predicted velocities. When the variation in predicted pose is small, the maneuver prediction is highly accurate, whereas larger variations are observed when the prediction accuracy is low.
                \item TS$1$ shows less variation compared to TS$\infty$, indicating a tendency for the predicted particles to concentrate more around the mean value. 
            \end{itemize}
                
        \subsection{Application to a simulator for evaluating heading-keeping PD control}\label{subsec:ctrl}
            This section demonstrates how the proposed prediction method of maneuvering motion functions as a simulator for evaluating ship maneuvering systems or control algorithms. In this paper, as a simple example, we present the results of evaluating the performance based on the PD control gains for heading-keeping. Here, a set of model obtained using the simulation data described in \Cref{subsec:sim_pred} is used, and the simulation environment based on the MMG model described in \Cref{subsec:sim_dataset} is used as the true system.
            
            Specifically, a $100.0 \ \mathrm{(s)}$ maneuvering simulation was conducted with an initial velocity of $u=0.5 \ \mathrm{(m/s)}, v_{\mathrm{m}}=0.0 \ \mathrm{(m/s)}, r=0.0 \ \mathrm{(rad/s)}$, a wind speed of $U_{\mathrm{T}}=0.0$, and a bow thruster rotational speed of $n_{\mathrm{BT}}=0 \ \mathrm{(rps)}$. The rudder angle is determined as follows:
            \begin{equation}
                \delta_{\mathrm{P}} = \delta_{\mathrm{S}} = - K_{\mathrm{P}} \left(\psi - \bar{\psi}\right) - K_{\mathrm{D}} r \enspace,
                \label{eq:PDcontroller}
            \end{equation}
            where $K_{\mathrm{P}}$ and $K_{\mathrm{D}}$ are gains, and $\bar{\psi}=\pi/2$ is the indicative heading angle. Note that during the simulation, the rudder angle is determined using the predicted values of each particle from \Cref{eq:PDcontroller}. The performance of the PD gains for a given particle trajectory is defined by the following metric:
            \begin{equation}
                \mathcal{L}_{\text{PD},p}\left( K_{\mathrm{P}}, K_{\mathrm{D}} \right) = \int_{t=0}^{100.0} \left\{\left|\psi_{p}\left(t\right) - \bar{\psi}\right| + \left|r_{p}\left(t\right)\right|\right\} \mathrm{d}t \enspace.
                \label{eq:PDscore}
            \end{equation}
            The performance value of the PD gain in the true system is then denoted as $\mathcal{L}_{\text{PD,true}}$.
    
\begin{figure}[t]
                \centering
                \includegraphics[width=\hsize, trim=0 3 0 0, clip]{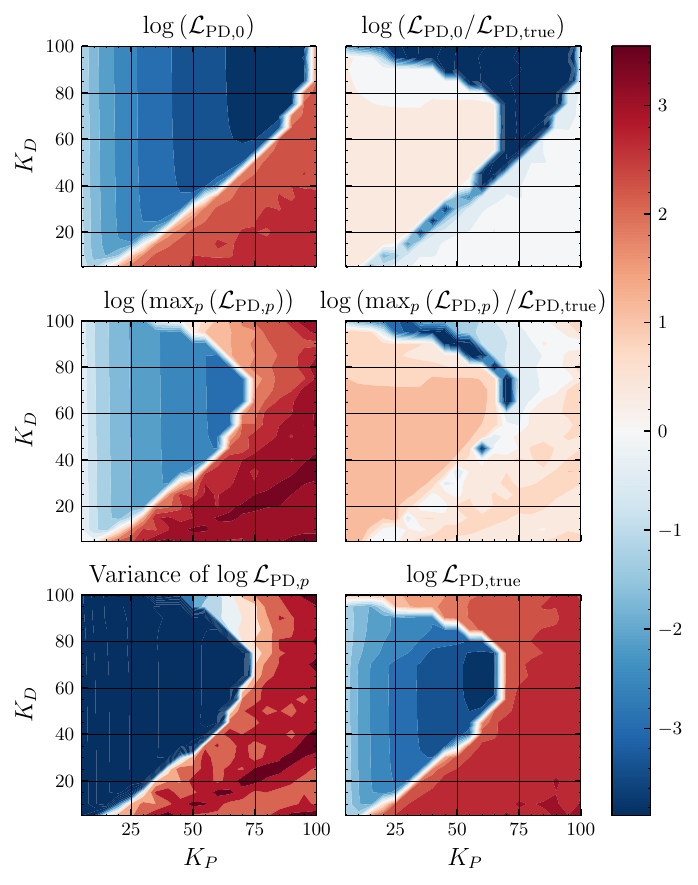}
                \caption{Evaluation results for PD gain performance.}
                \label{fig:compPD_cont}
            \end{figure}

\begin{figure*}[t]
                \centering
                \begin{minipage}[b]{\linewidth}
                    \centering
                    \begin{minipage}[b]{0.32\linewidth}
                        \centering
                        \includegraphics[width=0.95\hsize, trim=10 10 10 10, clip]{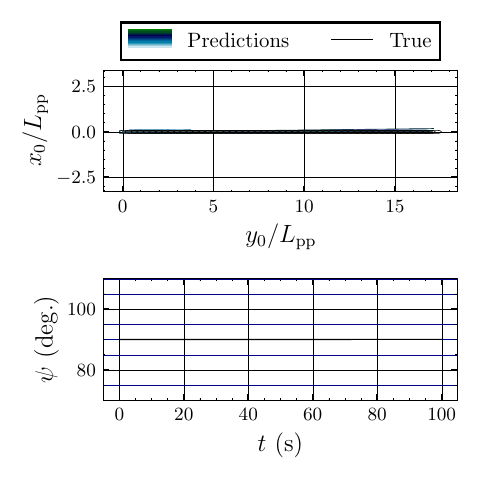}
                        \subcaption{Ship pose.}
                        \label{fig:P60.0_D60.0.csv_psi}
                    \end{minipage}
                    \begin{minipage}[b]{0.32\linewidth}
                        \centering
                        \includegraphics[width=0.95\hsize, trim=10 10 10 10, clip]{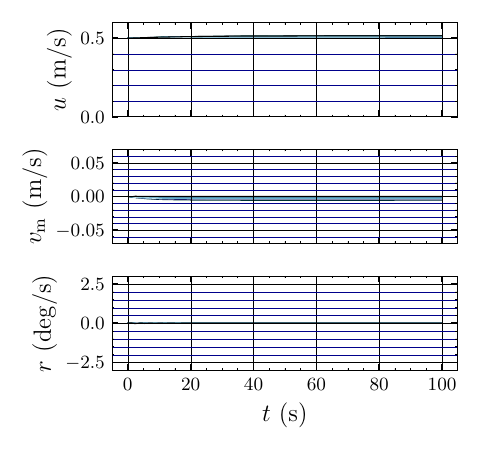}
                        \subcaption{Ship velocity.}
                        \label{fig:P60.0_D60.0.csv_nu}
                    \end{minipage}
                    \begin{minipage}[b]{0.32\linewidth}
                        \centering
                        \includegraphics[width=0.95\hsize, trim=10 10 10 10, clip]{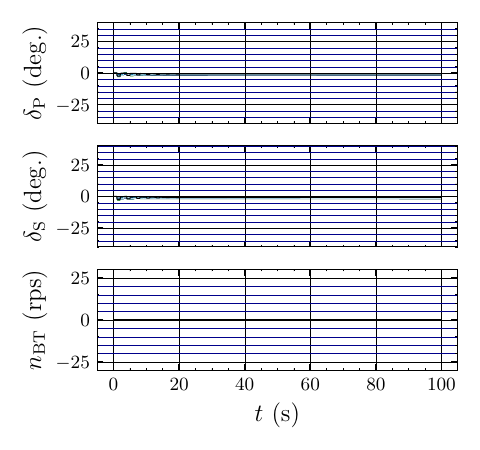}
                        \subcaption{Actuator state.}
                        \label{fig:P60.0_D60.0.csv_u}
                    \end{minipage}
                \end{minipage}
                \caption{Evaluation results of heading-keeping PD control using $\left(K_{\mathrm{P}}, K_{\mathrm{D}}\right)=\left(60, 60\right)$.}
                \label{fig:P60.0_D60.0}
            \end{figure*}

\begin{figure*}[t]
                \centering
                \begin{minipage}[b]{\linewidth}
                    \centering
                    \begin{minipage}[b]{0.32\linewidth}
                        \centering
                        \includegraphics[width=0.95\hsize, trim=10 10 10 10, clip]{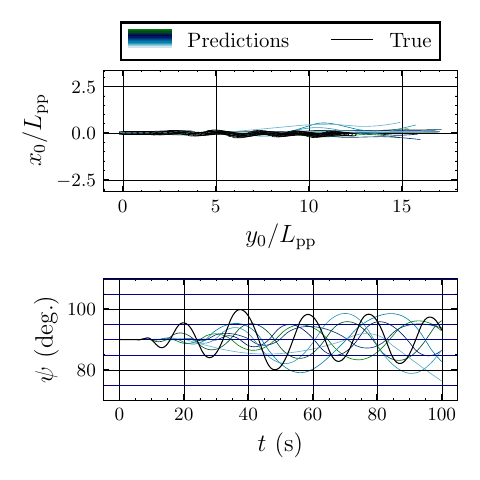}
                        \subcaption{Ship pose.}
                        \label{fig:P90.0_D90.0.csv_psi}
                    \end{minipage}
                    \begin{minipage}[b]{0.32\linewidth}
                        \centering
                        \includegraphics[width=0.95\hsize, trim=10 10 10 10, clip]{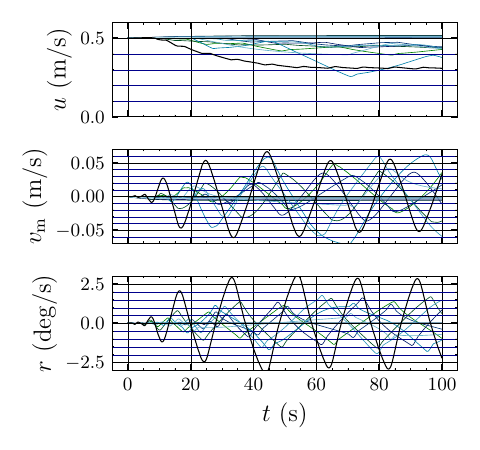}
                        \subcaption{Ship velocity.}
                        \label{fig:P90.0_D90.0.csv_nu}
                    \end{minipage}
                    \begin{minipage}[b]{0.32\linewidth}
                        \centering
                        \includegraphics[width=0.95\hsize, trim=10 10 10 10, clip]{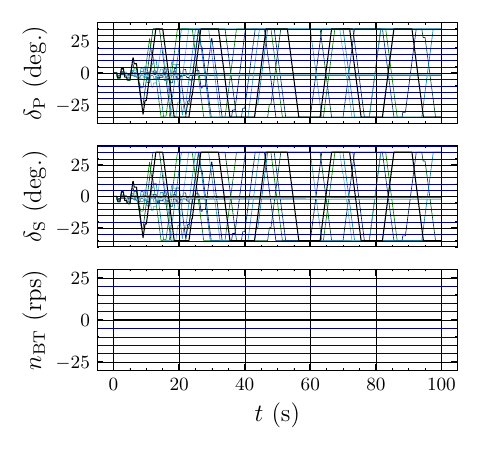}
                        \subcaption{Actuator state.}
                        \label{fig:P90.0_D90.0.csv_u}
                    \end{minipage}
                \end{minipage}
                \caption{Evaluation results of heading-keeping PD control using $\left(K_{\mathrm{P}}, K_{\mathrm{D}}\right)=\left(90, 90\right)$.}
                \label{fig:P90.0_D90.0}
            \end{figure*}

            The PD gains were evaluated for all combinations of $\left(K_{\mathrm{P}}, K_{\mathrm{D}}\right) \in \left\{5, 10, \cdots, 100\right\}\times\left\{5, 10, \cdots, 100\right\}$, and the results of the obtained scores are shown in \Cref{fig:compPD_cont}. From these results, the following observations can be drawn:
            \begin{itemize}
                \item The performance of the PD gains evaluated using a single maneuvering model tends to be overestimated compared to the performance in the true system, especially in the region where both $\left(K_{\mathrm{P}}, K_{\mathrm{D}}\right)$ are large.
                \item By considering the worst-performing predicted particle in terms of PD gain performance, it is possible to reduce the overestimation of performance compared to the true system.
            \end{itemize}
            Additionally, \Cref{fig:P60.0_D60.0,fig:P90.0_D90.0} show the simulation results for $\left(K_{\mathrm{P}}, K_{\mathrm{D}}\right) = \left(60, 60\right)$ and $\left(90, 90\right)$. From these results, the following observations can be made:
            \begin{itemize}
                \item When the PD gains function properly for heading-keeping control, the rudder angle and yaw rate show minimal variation, and the system does not transition into states not included in $\mathcal{D}_{\text{Train-B}}$. In such cases, the PD gain performance can be evaluated with high accuracy.
                \item When the PD gains do not function appropriately for heading-keeping control, the rudder angle and yaw rate exhibit significant fluctuations, and the system transitions into states not included in $\mathcal{D}_{\text{Train-B}}$. In these cases, although the performance evaluation accuracy is lower, the results vary significantly across predicted particles.
            \end{itemize}
            
    \section{Experiments using navigation data of a full-scale ship}\label{sec:nav_exp}
        This section presents the results of numerical experiments on the prediction accuracy of maneuvering motion using port navigation data of a full-scale ship. The dataset used in the experiments is described in \Cref{subsec:nav_dataset}. Subsequently, the prediction accuracy and uncertainty prediction results of the proposed method is demonstrated in \Cref{subsec:nav_pred}.
        
        \subsection{Dataset}\label{subsec:nav_dataset}
            This section describes the port navigation data for a full-scale ship. The dataset consists of the actual operational data of a full-scale ship of about 150 meters recorded over several months, which included the ship's state, control inputs, and apparent wind speed and direction observed at $1 \ \mathrm{(Hz)}$. In particular, data corresponding to maneuvers from entering port to berthing at the three ports, referred to as Port A, Port C, and Port D by Mwange et al. \cite{mwange2024}. The training and test datasets of the navigation data are selected as shown in \Cref{tab:test_full_dataset}.
            Here, the full-scale vessel is equipped with a controllable pitch bow thruster and the thrust is controlled by the pitch angle. The command current value of the pitch angle is recorded in the operational data and is denoted as $i_{\mathrm{BT}}$ and used instead of the bow thruster speed $n_{\mathrm{BT}}$. For details on the measurement method and statistical properties of these navigation data, please refer to the previous literature \cite{Miyauchi2022data,mwange2024}.
            
            \begin{table}[t]
                \centering
                \caption{Datasets of port navigation of full-scale ship}
                \begin{tabular}{lll}\toprule
                    Dataset & Maneuvers & Total Duration \\ \midrule
                    $\mathcal{D}_{\text{Train-Full-B}}$ & $\mathrm{B}_{1}, \cdots, \mathrm{B}_{27}$  & $32180 \ \mathrm{(s)}$ \\
                    $\mathcal{D}_{\text{Test-Full-B}}$ & $\mathrm{B}_{28}, \cdots, \mathrm{B}_{41}$ & $16230 \ \mathrm{(s)}$ \\
                    \bottomrule
                \end{tabular}
                \label{tab:test_full_dataset}
            \end{table}
                
        \subsection{Maneuvering motion prediction and its uncertainty}\label{subsec:nav_pred}
            This section presents the prediction results for maneuvering motion using the proposed method. The number of ensembles was set to $M=15$, and the hyperparameters shown in \Cref{tab:hyperparam} were used for training. In the evaluation of trained models, as in \Cref{subsec:nav_pred}, the initial state was given from the test dataset, and $P$ predicted values of the state variables were computed using the proposed method. Here, $P=100$ and $K=100$.
            
\begin{figure}[t]
                \centering
                \includegraphics[width=0.625\linewidth, trim=5 10 5 10, clip]{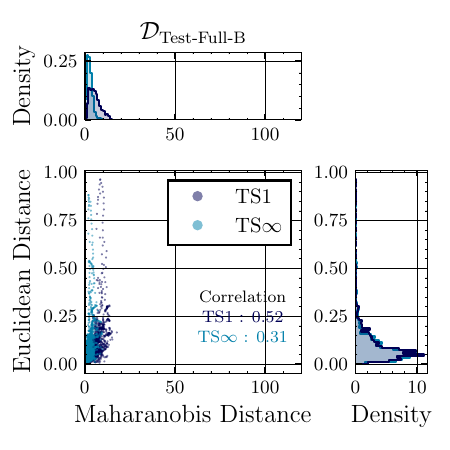}
                \caption{Distribution of Mahalanobis distance and Euclidean distance at $K=100$, $M=15$.}
                \label{fig:compts_dist_full}
            \end{figure}

            \Cref{fig:compts_dist_full} illustrates the relationship between the Euclidean distance between the mean of the predictive particles and the observed values, as well as the Mahalanobis distance between the set of predictive particles and the observed values, for the test dataset $\mathcal{D}_{\text{Test-Full-B}}$. The following insights can be drawn from these results:
            \begin{itemize}
                \item For the operational data, the correlation coefficient between the Euclidean and Mahalanobis distances is lower for TS$\infty$ than for TS$1$. Additionally, the magnitude of the Mahalanobis distance does not significantly differ from that observed in the simulation.
                \item Although the majority of Euclidean distance values are below $0.25$, some reach nearly $1.0$.
            \end{itemize}
    
\begin{figure*}[t]
                \centering
                \begin{minipage}[b]{0.32\linewidth}
                        \centering
                        \includegraphics[width=0.95\hsize, trim=8 8 8 8, clip]{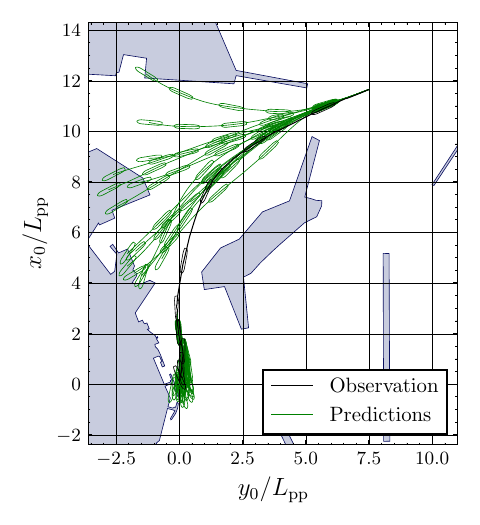}
                        \subcaption{Ship pose.}
                        \label{fig:TSinfty_kanda20220509in_x0y0}
                \end{minipage}
                \begin{minipage}[b]{0.66\linewidth}
                        \centering
                        \includegraphics[width=0.48\hsize, trim=8 8 8 8, clip]{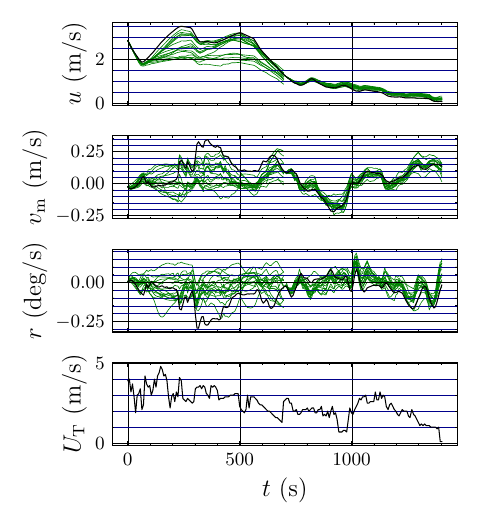}
                        \includegraphics[width=0.48\hsize, trim=8 8 8 8, clip]{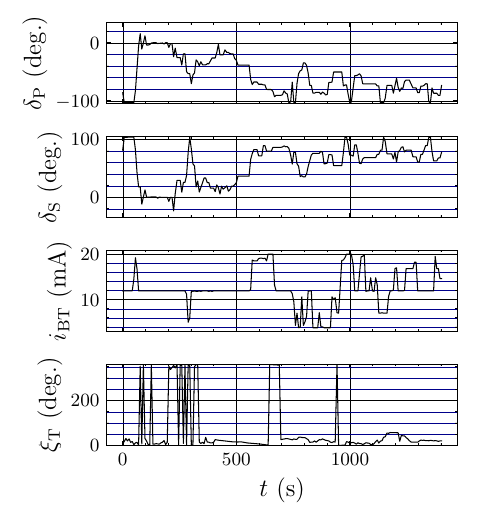}
                        \subcaption{Ship velocities, actuator states, and wind states.}
                        \label{fig:TSinfty_kanda20220509in_nu_UT}
                \end{minipage}
                \caption{Prediction of a berthing maneuver using TS$\infty$. Note that it is initialized using the observed value at $t=0.0 \ \mathrm{(s)}$ and $t=700.0 \ \mathrm{(s)}$.}
                \label{fig:TSinfty_kanda20220509in}
            \end{figure*}

\begin{figure*}[t]
                \centering
                \begin{minipage}[b]{0.32\linewidth}
                        \centering
                        \includegraphics[width=0.95\hsize, trim=8 8 8 8, clip]{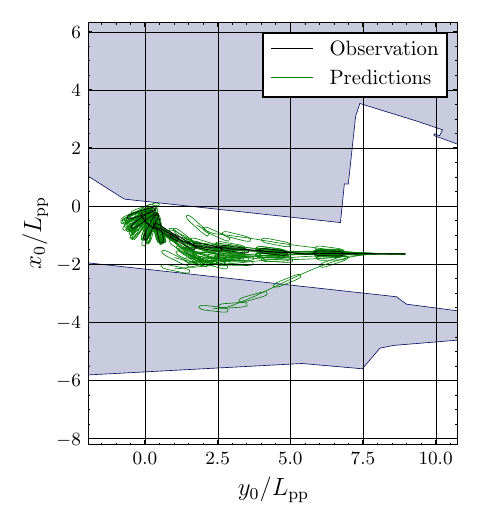}
                        \subcaption{Ship pose.}
                        \label{fig:TSinfty_sendai20220415in_x0y0}
                \end{minipage}
                \begin{minipage}[b]{0.66\linewidth}
                        \centering
                        \includegraphics[width=0.48\hsize, trim=8 8 8 8, clip]{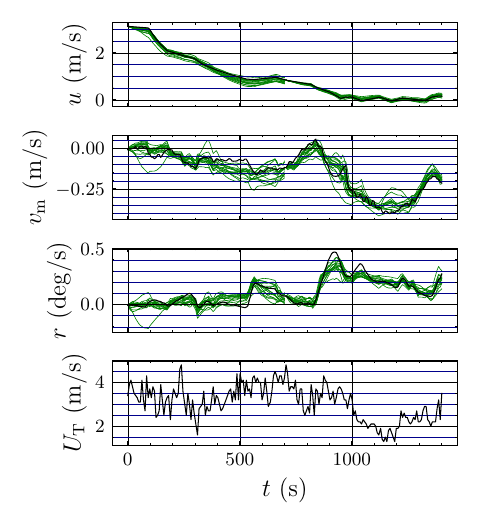}
                        \includegraphics[width=0.48\hsize, trim=8 8 8 8, clip]{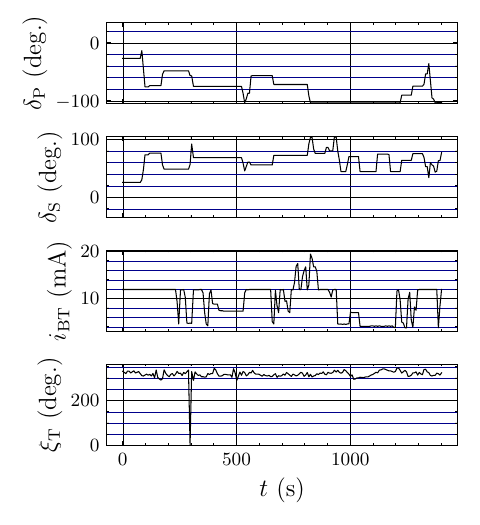}
                        \subcaption{Ship velocities, actuator states, and wind states.}
                        \label{fig:TSinfty_sendai20220415in_nu_UT}
                \end{minipage}
                \caption{Prediction of a berthing maneuver using TS$\infty$. Note that it is initialized using the observed value at $t=0.0 \ \mathrm{(s)}$ and $t=700.0 \ \mathrm{(s)}$.}
                \label{fig:TSinfty_sendai20220415in}
            \end{figure*}

            The prediction results of berthing maneuvers at Port A and Port C are shown in \Cref{fig:TSinfty_kanda20220509in,fig:TSinfty_sendai20220415in}. Notably, \Cref{fig:TSinfty_kanda20220509in} represents a case where the Euclidean distance reached nearly $1.0$. From these results, the following observations can be made:
            \begin{itemize}
                \item Between $100 \ \mathrm{(s)}$ and $700 \ \mathrm{(s)}$ in \Cref{fig:TSinfty_kanda20220509in}, the prediction accuracy is low for the surge velocity $u$, sway velocity $v_{\mathrm{m}}$, and yaw rate $r$. While the variance of predictive particles is high, there are segments where the bias relative to the observed values is larger than the variance.
                \item Conversely, except for the interval from $100 \ \mathrm{(s)}$ to $700 \ \mathrm{(s)}$ in \Cref{fig:TSinfty_kanda20220509in}, the proposed method demonstrates a high level of accuracy in predicting the berthing maneuver with minimal variance in the predictive particles.
            \end{itemize}
            
    \section{Discussion}\label{sec:discussion}
        In \Cref{sec:sim_exp,sec:nav_exp}, we presented the results of numerical experiments on the prediction accuracy of maneuvering motions. These findings from these experiments can be summarized as follows:
        \begin{itemize}
            \item The introduction of the FNN $\boldsymbol{\phi}^{\prime}_{\boldsymbol{\theta}^{\prime}}$ for initial state estimation improved the fitting accuracy to the training data.  
            \item It was found that, for test dataset of berthing maneuvers, which has a distribution similar to the training dataset, relatively high accuracy and low variance predictions were achieved in the simulation data and navigation data. However, in the navigation data, it was observed that, although high accuracy and low variance predictions is possible, the bias between predicted particles and observations could be relatively large compared to their variance in the navigation data.
            \item In the proposed method, increasing the ensemble size did not reduce the bias between the predicted particle set and the true value. However, when the bias was small, the variance was also small, and when the bias was large, the variance tended to increase. Furthermore, this trend was more pronounced when using TS$\infty$ compared to TS$1$.
            \item In the evaluation of the gain performance for heading-keeping PD control, it was found that the proposed probabilistic prediction method of maneuvering motions could reduce the likelihood of overestimating performance compared to simulations using a single FNN-based maneuvering model by considering the worst-case score. However, if motion states suitable from the perspective of control purposes were included in the training data, the prediction accuracy of maneuvering would be high, meaning that even when considering the worst-case performance, the evaluation accuracy did not significantly deteriorate.
        \end{itemize}
    
\begin{figure}[t]
            \centering
            \includegraphics[width=\hsize, trim=10 10 10 10, clip]{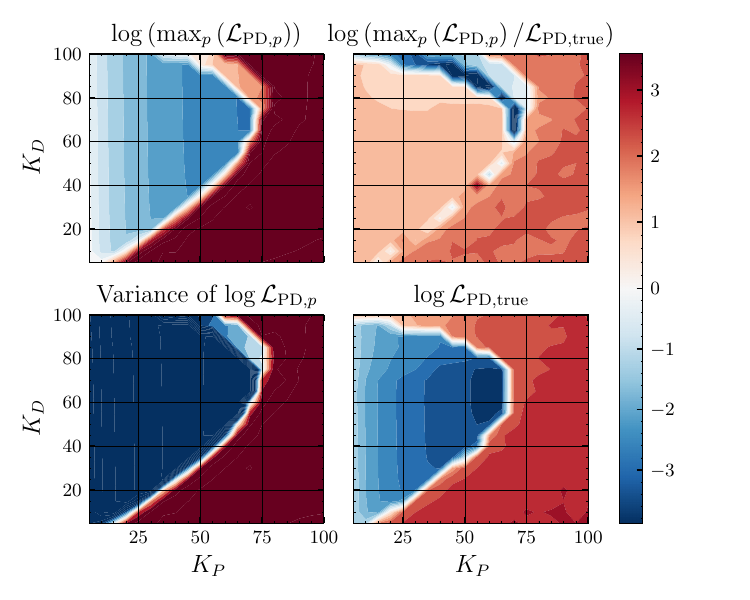}
            \caption{Maximum score for PD gain performance using a set of maneuvering models with $M=75$.}
            \label{fig:compPD_75_cont}
        \end{figure}

        Nevertheless, as seen in \Cref{fig:compPD_cont}, the proposed method did not completely eliminate the possibility of overestimating the gain performance in heading-keeping PD control. Here, the results for a larger ensemble size ($M=75$) are presented in \Cref{fig:compPD_75_cont}. These results indicate that considering a greater number of possibilities reduces the likelihood of overestimation. However, it was still not possible to entirely eliminate overestimation on the boundary where performance rapidly deteriorates with small changes in gain. 
        In heading-keeping PD control, it is expected that the evaluation results will vary significantly depending on whether the system transitions to a state where stability is lost. In other words, in the problem setting of this study, even minor prediction errors could lead to substantial differences in the evaluation outcomes. Therefore, it is possible that overestimation at the boundary is caused by uncertainty due to measurement errors included in the training data. Thus, while increasing the ensemble size can reduce the likelihood of overestimation, it is important to note that it does not completely eliminate the possibility of overestimation. 
    
        Furthermore, in the navigation data, a result where the prediction accuracy decreased was also shown. This performance degradation could have been caused by observation errors, the effects of unobserved variables such as waves and currents, the effects of draft changes, and so on. It should be noted that the proposed method can capture epistemic uncertainty, but cannot represent aleatoric uncertainty due to partial observability or disturbances.
    
        Despite these limitations, the proposed method enables the probabilistic prediction of ship maneuvering motion by capturing epistemic uncertainty. Consequently, the proposed method not only reduces the possibility of overestimating performance of maneuvering systems but also facilitates the design of maneuvering control algorithms that incorporate robust and risk-averse control. Although further validation is required, the proposed method shows potential for the proposed method to be applied to full-scale vessels. Therefore, a future work is to verify the effectiveness of the proposed method as a simulator for the design and evaluation of ship maneuvering systems.
        
    \section{Conclusion}\label{sec:conclusion}
        This paper proposes an ensemble learning method to capture epistemic uncertainty arising from insufficient or unevenly distributed data within a non-parametric SI using ANNs for ship maneuvering motion prediction. Furthermore, a probabilistic motion prediction method based on the obtained ensemble of models is introduced. The effectiveness of the proposed method is demonstrated by examining the relationship between prediction accuracy and uncertainty in maneuvering motion predictions, utilizing simulation data from a model ship and navigation data from a full-scale ship.
                
        The revealed findings are as follows:        
        \begin{itemize}
            \item Introducing an FNN for initial state estimation improves fitting accuracy of velocity components for training dataset.
            \item The proposed method tends to exhibit low variance when the bias between the ensemble of prediction particles and the true value is small, and high variance when the bias is large. This trend is more significant when using TS$\infty$ as a trajectory sampling method compared to TS$1$.
            \item Given training data of berthing maneuvers, the proposed method can produce highly accurate and low-variance predictions of berthing maneuvers, whose state distributions are close to the training data, and it is able to make low-accuracy but high-variance predictions of Zigzag and Turning maneuvers whose state distributions are not close to the training data.
            \item As a simulation environment for evaluating maneuvering systems, considering the worst score within the proposed probabilistic motion prediction results helps mitigate the possibility of overestimating performance relative to the true system.
        \end{itemize}
        Therefore, it is demonstrated that the proposed method is a non-parametric SI capable of probabilistically predicting ship maneuvering motion by capturing epistemic uncertainty.

    \section*{Acknowledgements}
        This study was supported by a Grant-in-Aid for Scientific Research from the Japan Society for the Promotion of Science (JSPS KAKENHI Grant \#22H01701 and \#23KJ1432). 

\bibliographystyle{spphys}       
\bibliography{reference.bib}   

\appendix

    \section{Simulation models}\label{sec:app_simulator}
        \subsection{MMG model}\label{subsec:mmg_model}
            The MMG model used in this study is described in detail. The three-degree-of-freedom equations of motion, commonly applied in standard MMG models, are defined as follows:
            \begin{equation}
                \left\{ \,
                  \begin{aligned}
                    \left(m+m_x\right) \dot{u}-\left(m+m_y\right) v_m r-x_G m r^2 & =X \\
                    \left(m+m_y\right) \dot{v}_m+\left(m+m_x\right) u r+x_G m \dot{r} & =Y \\
                    \left(I_{z z}+J_{z z}+x_G^2 m\right) \dot{r}+x_G m\left(\dot{v}_m+u r\right) & =N \enspace,
                  \end{aligned}
                \right.
                \label{eq:mmg_model}
            \end{equation}
            where $X$, $Y$, and $N$ are the forces and moments acting on the ship, decomposed as follows:
            \begin{equation}
                \left\{ \,
                  \begin{aligned}
                    X & =X_{\mathrm{H}}+X_{\mathrm{P}}+X_{\mathrm{R}}+X_{\mathrm{BT}}+X_{\mathrm{A}} \\
                    Y & =Y_{\mathrm{H}}+Y_{\mathrm{P}}+Y_{\mathrm{R}}+Y_{\mathrm{BT}}+Y_{\mathrm{A}} \\
                    N & =N_{\mathrm{H}}+N_{\mathrm{P}}+N_{\mathrm{R}}+n_{\mathrm{BT}}+N_{\mathrm{A}} \enspace.
                  \end{aligned}
                \right.
                \label{eq:mmg_force_model}
            \end{equation}
            Here, the subscripts H, P, R, BT, and A denote forces attributable to the hull, propeller, rudder, bow thruster, and apparent wind, respectively. The hull hydrodynamics are modeled using Yoshimura's unified model \cite{Yoshimura2009}. The resistance coefficients and linear hydrodynamic derivatives were determined through captive model testing, while the remaining coefficients were estimated using empirical formulas from Yoshimura \cite{Yoshimura2009}. For rudder and propeller forces, Kang's model \cite{kang2008} was employed. The propeller thrust coefficient was derived from open-water propeller tests, and the other coefficients were based on the VLCC values provided by Kang \cite{kang2008}.
    
            The thruster force was modeled based on Kobayashi's model \cite{Kobayashi1988} as follows:
            \begin{equation}
                \left\{ \,
                \begin{aligned}
                    X_{\mathrm{BT}} & =0 \\
                    Y_{\mathrm{BT}} & =\left(1+a_{\mathrm{YSB1}}+a_{\mathrm{YSB2}} \cdot \mathrm{Fr}+a_{\mathrm{YSB3}} \cdot \mathrm{Fr}^2\right) \cdot T_{B T} \\
                    N_{\mathrm{BT}} & =\left(1+a_{\mathrm{NSB1}}+a_{\mathrm{NSB2}} \cdot \mathrm{Fr}+a_{\mathrm{NSB3}} \cdot \mathrm{Fr}^2\right) \cdot T_{\mathrm{BT}} \cdot x_{\mathrm{BT}} \\
                    T_{\mathrm{BT}} & =\rho D_{\mathrm{BT}}^4 n_{\mathrm{BT}}^2 K_{\mathrm{T,BT}} \enspace.
                \end{aligned}
                \right.
                \label{eq:kobayashi_model}
            \end{equation}
            Here, $a_{\mathrm{YSB1}}$, $a_{\mathrm{YSB2}}$, $a_{\mathrm{YSB3}}$, $a_{\mathrm{NSB1}}$, $a_{\mathrm{NSB2}}$, $a_{\mathrm{NSB3}}$ are the coefficients of a quadratic function of the Froude number, representing the relationship between thruster force reduction and ship speed, and $x_{\mathrm{BT}}$ is the longitudinal position of the side thruster. These coefficients were determined through captive model testing, while the thrust coefficient $K_{\mathrm{T,BT}}$ was obtained experimentally.
            
            For wind forces, Fujiwara's regression model \cite{Fujiwara1998}, represented by the following equations, was used:
            \begin{equation}
                \left\{ \,
                \begin{aligned}
                    X_{\mathrm{A}} & =(1 / 2) \rho_{\mathrm{A}} U_{\mathrm{A}}^2 A_{\mathrm{T}} \cdot C_X \\
                    Y_{\mathrm{A}} & =(1 / 2) \rho_{\mathrm{A}} U_{\mathrm{A}}^2 A_{\mathrm{L}} \cdot C_Y \\
                    N_{\mathrm{A}} & =(1 / 2) \rho_{\mathrm{A}} U_{\mathrm{A}}^2 A_{\mathrm{L}} L_{{\mathrm{OA}}} \cdot C_N  \enspace,
                \end{aligned}
                \right.
                \label{eq:fujiwara_model}
            \end{equation}
            where
            \begin{equation}
                \left\{ \,
                \begin{aligned}
                    & C_X=X_0+X_1 \cos \left(2 \pi-\gamma_{\mathrm{A}}\right)+X_3 \cos 3\left(2 \pi-\gamma_{\mathrm{A}}\right)+X_5 \cos 5\left(2 \pi-\gamma_{\mathrm{A}}\right) \\
                    & C_Y=Y_1 \sin \left(2 \pi-\gamma_{\mathrm{A}}\right)+Y_3 \sin 3\left(2 \pi-\gamma_{\mathrm{A}}\right)+Y_5 \sin 5\left(2 \pi-\gamma_{\mathrm{A}}\right) \\
                    & C_N=N_1 \sin \left(2 \pi-\gamma_{\mathrm{A}}\right)+N_2 \sin 2\left(2 \pi-\gamma_{\mathrm{A}}\right)+N_3 \sin 3\left(2 \pi-\gamma_{\mathrm{A}}\right) \enspace.
                \end{aligned}
                \right.
                \label{eq:fujiwara_coeff}
            \end{equation}
            Here, $\rho_{\mathrm{A}}$ denotes the density of air, and $A_{\mathrm{T}}$, $A_{\mathrm{L}}$, and $L_{{\mathrm{OA}}}$ represent the transverse projected area, lateral projected area, and overall length of the ship, respectively. The coefficients $X_i,Y_i,N_i$ are wind pressure coefficients derived from regression formulas based on wind tunnel tests using scale models, with the geometric parameters of the ship as explanatory variables.
            
        \subsection{Actuator response model}\label{subsec:act_model}
            The dynamics of the actuator state variables are described as follows. The rotational speeds of the propeller, thrusters, and rudder angle cannot change instantaneously and exhibit some delay relative to their command values. Therefore, in the simulator, constraints were applied to ensure that the rate of change of the actuator state variables remains constant. Letting $y \in \boldsymbol{u}$ represent the components of the actuator state variables and $r \in \boldsymbol{a}$ denote the corresponding command values, the response characteristics of the actuators can be modeled as follows:
            \begin{equation}
                \dot{y} = K f_{\mathrm{step}}\left(r-y\right) \enspace.
                \label{eq:1dfilter_discrete}
            \end{equation}
            Here, $K$ is a constant representing the magnitude of the rate of change, and $f_{\mathrm{step}}$ is a step function with a defined slope, as specified below:
            \begin{equation}
                f_{\mathrm{step}}\left(y\right)=
                    \left\{\begin{array}{ll}
                        1, & \left(\epsilon \leq y\right) \\
                        y / \epsilon, & \left(\epsilon < y < \epsilon \right) \\
                        -1, & \left(y \leq \epsilon\right) \enspace,
                    \end{array}\right. 
                \label{eq:step_slope}
            \end{equation}
            where $\epsilon$ is a constant. The slope of the step function is introduced to prevent oscillations in the actuator state variables during numerical simulations using the Euler method, with $\epsilon$ set equal to the simulation time step. The rate of change $K$ for $\delta_{\mathrm{P}}$, $\delta_{\mathrm{S}}$, and $\delta_{\mathrm{BT}}$ is $20 \ \mathrm{(deg./s)}$, $20 \ \mathrm{(deg./s)}$, and $20 \ \mathrm{(rps/s)}$, respectively.
            
        \subsection{Stochastic wind process}\label{subsec:wind_model}
            The wind process was generated using the method proposed by Maki et al. \cite{Maki2022, Maki2024}. In this method, it is assumed that the processes of wind speed and wind direction are represented by the following one-dimensional filter equations:
            \begin{equation}
                \left\{ \begin{array}{l}
                    \mathrm{d} U_{\mathrm{T}} =\alpha_{U}\left(U_{\mathrm{T}}-\bar{U}_{\mathrm{T}}\right) \mathrm{d} t+\sigma_{U} \mathrm{d} W \\
                    \mathrm{d} \gamma_{\mathrm{T}} =\alpha_{\gamma}\left(\gamma_{\mathrm{T}}-\bar{\gamma}_{\mathrm{T}}\right) \mathrm{d} t+\sigma_{\gamma} \mathrm{d} W \enspace.
                \end{array} \right.
                \label{eq:1dfilter}
            \end{equation}
            Here, $\alpha_{U}<0,\alpha_{\gamma}<0,\sigma_{U},\sigma_{\gamma}$are filter coefficients, $\bar{U}_{\mathrm{T}},\bar{\gamma}_{\mathrm{T}}$ denote the mean values of wind speed and wind direction, and $\mathrm{d} W$ represents the increment of a Wiener process. The filter coefficient related to wind speed is determined using the wind speed spectrum approximated by the Hino spectrum, while the filter coefficient for wind direction is determined using a regression formula proposed by Kuwashima et al., which relates the standard deviation of wind direction fluctuations to the mean wind speed.

\end{document}